\documentclass[twocolumn,english,pra,reprint]{revtex4-2}
\usepackage[utf8]{inputenc}
\usepackage{color}
\usepackage{babel}
\usepackage{verbatim}
\usepackage{amsmath}
\usepackage{graphicx}
\usepackage{xargs}[2008/03/08]
\usepackage[unicode=true,pdfusetitle,
 bookmarks=true,bookmarksnumbered=false,bookmarksopen=false,
 breaklinks=false,pdfborder={0 0 1},backref=false,colorlinks=true]
 {hyperref}

\makeatletter
\usepackage{orcidlink}
\usepackage[caption=false]{subfig} 
\usepackage{ragged2e} 
\DeclareCaptionJustification{justified}{\justifying}

\@ifundefined{showcaptionsetup}{}{%
 \PassOptionsToPackage{caption=false}{subfig}}
\usepackage{subfig}
\makeatother

\begin{document}
\title{Optimization of probabilistic quantum search algorithm with \emph{a
priori} information}
\author{Yutong Huang$^{1}$}
\author{Shengshi Pang$^{1,2}$\orcidlink{0000-0002-6351-539X}}
\email{pangshsh@mail.sysu.edu.cn}

\affiliation{$^{1}$School of Physics, Sun Yat-sen University, Guangzhou, Guangdong
510275, China}
\affiliation{$^{2}$Hefei National Laboratory, University of Science and Technology
of China, Hefei 230088, China}
\begin{abstract}
A quantum computer encodes information in quantum states and runs
quantum algorithms to surpass the classical counterparts by exploiting
quantum superposition and quantum correlation. Grover's quantum search
algorithm is a typical quantum algorithm that proves the superiority
of quantum computing over classical computing. It has a quadratic
reduction in the query complexity of database search, and is known
to be optimal when no \emph{a priori} information about the elements
of the database is provided. In this work, we consider a probabilistic
Grover search algorithm allowing nonzero probability of failure for
a database with a general \emph{a priori} probability distribution
of the elements, and minimize the number of oracle calls by optimizing
the initial state of the quantum system and the reflection axis of
the diffusion operator. The initial state and the reflection axis
are allowed to not coincide, and thus the quantum search algorithm
rotates the quantum system in a three-dimensional subspace spanned
by the initial state, the reflection axis and the search target state
in general. The number of oracle calls is minimized by a variational
method, and formal results are obtained with the assumption of low
failure probability. The results show that for a nonuniform \emph{a
priori} distribution of the database elements, the number of oracle
calls can be significantly reduced given a small decrease in the success
probability of the quantum search algorithm, leading to a lower average
query complexity to find the solution of the search problem. The results
are applied to a simple but nontrivial database model with two-value
\emph{a priori} probabilities to show the power of the optimized quantum
search algorithm. The paper concludes with a discussion about the
generalization to higher-order results that allows for a larger failure
probability for the quantum search algorithm.

\newcommandx\ket[1][usedefault, addprefix=\global, 1=]{|#1\rangle}%
\newcommandx\bra[1][usedefault, addprefix=\global, 1=]{\langle#1|}%
\newcommandx\pro[1][usedefault, addprefix=\global, 1=]{|#1\rangle\langle#1|}%
\global\long\def\ord#1{\left.#1\right|_{\text{std}}}%
\newcommandx\pros[2][usedefault, addprefix=\global, 1=]{|#1\rangle\langle#2|}%
\global\long\def\exa#1{\left.#1\right|_{\text{exa}}}%
\global\long\def\arccsc{{\rm arccsc}}%
\global\long\def\arcsec{{\rm arcsec}}%
\end{abstract}
\maketitle

\section{Introduction}

Quantum computing has been expected to revolutionize the field of
computing since it was proposed \citep{benioff1980thecomputer,feynman1982simulating}.
It accelerates computing tasks by taking advantage of the nonclassicalities
of quantum systems such as quantum superposition and quantum correlation.
With the development of quantum computing, various quantum algorithms
have been proposed. The large number factorization algorithm proposed
by Peter W. Shor \citep{shor1994algorithms,shor1997polynomialtime}
has an exponential speedup compared to classical factorization algorithms,
and the quantum search algorithm proposed by Lov K. Grover \citep{grover1997quantum,grover1998quantum}
has a quadratic speedup in terms of the database size compared to
classical database search algorithms. More quantum algorithms have
been proposed in recent years, such as the variational quantum eigensolver
algorithm \citep{peruzzo2014avariational} and the quantum approximate
optimization algorithms \citep{farhi2014aquantum} for noisy intermediate-scale
quantum devices, the Harrow-Hassidim-Lloyd algorithm for linear systems
of equations applicable to quantum machine learning \citep{harrow2009quantum},
the boson sampling \citep{aaronson2013thecomputational} for photon
distribution in linear optics, etc.

For a classical database with $N$ elements, if a search task has
$M$ solutions, the classical query complexity of finding a solution
is usually of order $O(N/M)$%
. In quantum computing, Grover's search algorithm finds a solution
in a database by preparing the quantum system in an appropriate superposed
state of the computational basis and driving the system to approach
a target state with alternate oracle operations and specific diffusion
operations. After the evolution, a quantum projective measurement
is performed along the computational basis on the system to obtain
the target state. The query complexity required by Grover's search
algorithm is $O(\sqrt{N/M})$, which is a quadratic speedup compared
to a classical search algorithm. Because of this superiority, quantum
search algorithms have been found useful in various applications,
such as quantum dynamic programming \citep{ambainis2019quantum},
quantum random-walk search algorithm \citep{shenvi2003quantum}, the
preparation of Greenberger-Horne-Zeilinger states using algorithms
\citep{hao-sheng2000preparation}, etc.

While the quantum search algorithm has found wide applications in
vast areas, the algorithm itself can still be extended and improved
in various aspects. For example, the quantum partial search algorithm
\citep{grover2005ispartial,korepin2005optimization,korepin2006simplealgorithm,korepin2006questfor,choi2007quantum,giri2017areview,zhang2020depthoptimization}
decomposes the database into smaller blocks and searches for the block
which the target state belongs to instead of finding the exact location
of the target state. Another idea is to change the Grover iterations
of the quantum search algorithm. The core of Grover's quantum search
algorithm is the amplitude amplification technique \citep{Brassard2002,kwon2021quantum}
which increases the weight of the search target in the superposed
state of the quantum system by repetition of the Grover iteration
consisting of the oracle operation and a diffusion operation. %
{} It has been proposed that the diffusion operation in the Grover iteration
can be replaced by a two-dimensional phase rotation \citep{Hoyer2000}.
Extension of the two-dimensional phase rotation to a three-dimensional
rotation was subsequently proposed, and a phase matching condition
to realize the quantum search algorithm by phase rotation was found
\citep{long1999phasematching,long2002phasematching,nielsen2012quantum}.
This phase-matching condition has been verified theoretically \citep{biham2000analysis,Hoyer2000},
and realized by experiments on optical systems \citep{bhattacharya2002implementation,puentes2004optical}
and ion trap systems\citep{ivanov2008simpleimplementation}, etc.
An interesting zero-failure quantum search algorithm based on this
condition was subsequently proposed by Long \citep{long2001groveralgorithm},
and this algorithm has been applied in different quantum tasks, such
as quantum pattern recognition \citep{trugenberger2002quantum}, sliding
mode control of quantum systems \citep{dong2009sliding} and quantum
image compression \citep{chao-yang2006ahybrid}. Recently, an improved
two-parameter modified version \citep{roy2022deterministic} was proposed
to realize deterministic search without user control of the quantum
oracle. More variants of the quantum search algorithm have also been
proposed, such as the quantum search algorithm with continuous variables
\citep{pati2000quantum,heinrich2002quantum,roland2003quantumcircuit},
the fixed-point quantum search algorithm \citep{xiao2005errortolerance,yoder2014fixedpoint},
the Hamiltonian search algorithm \citep{roland2003quantumcircuit},
etc. A review of extended quantum search algorithms can be referred
to \citep{younes2013towards,morales2018variational,wan-ning2018useridentifying,plekhanov2022variational,mandal2023invariance}.
Moreover, the quantum search algorithm has been realized experimentally
on various physical systems, including photons \citep{kwiat2000grovers,walther2005experimental},
superconducting circuits \citep{dicarlo2009demonstration,roy2020programmable},
nuclear magnetic resonance \citep{jones1998implementation,ermakov2002experimental,jing-fu2003nmranalogue},
trapped ions \citep{brickman2005implementation,figgatt2017complete},
etc.

The lower bounds for the number of oracle calls of these algorithms
have been proven optimal \citep{boyer1999tightbounds,beals2001quantum,dohotaru2009exactquantum}.
In the optimality of the quadratic speedup of Grover's search algorithm,
a key ingredient is that all the elements of the database are equally
likely to be the solution of the search problem. If the elements of
the database are allowed to be unequally likely to be the search solution,
the result can be quite different, as it can be expected that preparing
the quantum system closer to the states with higher probabilities
to be the search target will be more beneficial to the search algorithm
\citep{sadowski2017quantum,dogra2018superposing,he2020quantum,ccalikyilmaz2022quantum}.
An intuitive example is that, if some of the database elements, e.g.,
$K$ of the $N$ elements, are known to have very low probabilities
(or even zero probabilities) to be the search target, one just needs
to prepare the quantum system in a superposition of the remaining
$N-K$ elements and flip the system around this state as well in the
Grover iterations, and then the query complexity will be reduced to
$O(\sqrt{N-K})$ rather than $O(\sqrt{N})$ at the cost of a low probability
to fail the search task. This inspires us to consider the following
question: if we know in advance the \emph{a priori} probabilities
of the elements in the database to be the search target, what is the
optimal performance of the quantum search algorithm obtained by exploiting
the \emph{a priori} probabilities of the database elements and allowing
the algorithm to succeed probabilistically?

The purpose of this work is to study the minimal query complexity
of the quantum search algorithm by optimizing the initial state of
the quantum system and the reflection axis of the diffusion operator
given the average success probability of the algorithm. The query
complexity is quantified by the number of oracle calls, and the minimization
of the query complexity is carried out by variation of the initial
state of the quantum system and the reflection axis of the diffusion
operator. The optimization conditions turn out to be highly nontrivial
and hard to solve in general. However, if the failure probability
is low, one can expect that the optimal initial state and the optimal
reflection axis should just deviate slightly from the uniformly superposed
state of all database elements as in the original Grover search algorithm.
This leads to a differential approach to solving the optimization
equations: by taking the differentiation of the optimization equations
as well as the normalization conditions for the initial state and
the reflection axis, one can establish a differential relation between
the success probability of the algorithm, the number of the oracle
calls, the optimal initial state of the system and the optimal reflection
axis of the diffusion operator. When the failure probability of the
algorithm is low, this differential relation can approximately tell
the reduction of the number of oracle calls in terms of the failure
probability of the quantum search algorithm.

In this work, based on the above idea, we obtain a formal second-order
differential relation between the failure probability of the algorithm
and the reduction in the number of oracle calls given a general \emph{a
priori} probability distribution of the database elements. An interesting
property of the result is that the reduction percentage of the number
of oracle calls is proportional to the square root of the failure
probability of the quantum search algorithm which is always much larger
than the failure probability when the latter is small, implying the
optimized probabilistic quantum search algorithm can decrease the
\emph{average} number of oracle calls when the success probability
of the algorithm is taken into account. The formal results are applied
to a simple but nontrivial database model where all the elements have
only two possible values for the \emph{a priori} probabilities, and
the reduction of the query complexity is analytically derived in terms
of the failure probability of the search algorithm and illustrated
in detail by numerical computation.

The paper is structured as follows. In Sec. \ref{sec:Preliminaries},
we give a brief overview for Grover's quantum search algorithm. In
Sec. \ref{sec:Optimization method}, the number of oracle calls is
minimized by the method of Lagrange multipliers given the \emph{a
priori} probabilities of the database elements and the success probability
of the search algorithm. The failure probability of the algorithm
is then assumed to be small, and the reduction of the number of oracle
calls is obtained in terms of the failure probability by differentiating
the optimization equations. Sec. \ref{sec:Analysis and Results} is
devoted to a simple database model with the \emph{a priori} probabilities
of the elements taken to be two valued. The paper finally concludes
in Sec. \ref{sec:Conclusion-and-outlook} with a summary of the work
and a discussion of the generalization to higher order results that
allow for a larger failure probability of the quantum search algorithm.

\section{Preliminaries\label{sec:Preliminaries}}

In this section, we briefly introduce the preliminaries of Grover's
quantum search algorithm \citep{grover1997quantum,grover1998quantum}
relevant to the current research. We focus on the case that the search
problem has only one solution throughout this paper.

\subsection{Procedures of Grover's search algorithm}

Suppose we have a database with $N$ elements where the probabilities
of the elements being the search target are the same, and we use the
computational basis of a quantum system to represent the elements
of the database. The quantum system is initially prepared in a uniform
superposed state,
\begin{equation}
\left|\psi_{0}\right\rangle =\sum_{i=1}^{N}\frac{1}{\sqrt{N}}\left|i\right\rangle .\label{eq:psi0}
\end{equation}

The solution of the search problem is recognized by a quantum oracle.
The quantum oracle can be regarded as a black box, the internal working
mechanism of which is not critical to the search algorithm, but can
perform a unitary transformation on the quantum system and mark up
the solution of the search task by shifting the phase of the target
state. In detail, the unitary transformation of the quantum oracle
can be written as
\begin{equation}
O=I-2\pro[t],
\end{equation}
where $\ket[t]$ is the target state and $I$ is the identity operator
on the $N$-dimensional Hilbert space of the system. The effect of
the oracle $O$ when it acts on a quantum state is that
\begin{equation}
O|t\rangle=-|t\rangle,\;O|i\rangle=|i\rangle,\;i\neq t.
\end{equation}
So, it flips the sign of the target state and leaves the basis states
other than the target state unchanged.

While the oracle can mark up the solution by changing the sign of
the target state, it cannot lead the quantum system to approach the
target state alone as it does not change the amplitude distribution
of different basis states in the superposed state of the quantum system.
In order to increase the amplitude of the target state in the superposed
state of the quantum system, the oracle operation needs to be followed
by another unitary transformation, usually called the Grover diffusion
operator, which reflects the state of the quantum system around the
uniformly superposed state $|\psi_{0}\rangle$,
\begin{equation}
D=2\pro[\psi_{0}]-I.
\end{equation}
The effect of the diffusion operator $D$ is to invert the amplitudes
of the basis states in the superposed state of the system around the
mean of all amplitudes. The combination of the oracle $O$ and the
diffusion operator $D$ is usually called the Grover operator or Grover
iteration, defined as
\begin{equation}
G=DO.
\end{equation}
It turns out that the Grover operator $G$ can boost the amplitude
of the target state in the superposed state of the system. So if one
repeats this procedure for a proper number of times, the quantum system
can finally approach the target state of the search problem with a
high fidelity.

Grover's search algorithm has an intuitive geometric interpretation.
In order to see how Grover's algorithm works in the geometric picture,
the initial state can be rewritten as
\begin{equation}
\begin{aligned}\begin{aligned}\ket[\psi_{0}]= & a_{t}\left|t\right\rangle +\sum_{i\neq t}^{N}a_{i}\left|i\right\rangle =\sin\theta\left|t\right\rangle +\cos\theta\left|t_{\perp}\right\rangle ,\end{aligned}
\end{aligned}
\end{equation}
where the state is decomposed to two states, one the target state
$|t\rangle$ and the other a uniformly superposed state in the remaining
$(N-1)$-dimensional subspace orthogonal to the target state $t$,
and
\begin{equation}
\sin\theta=\frac{1}{\sqrt{N}},\;\cos\theta=\sqrt{1-\frac{1}{N}}.
\end{equation}
It can be verified that after $j$ repetitions of the Grover iteration,
the initial state of the quantum system is transformed to
\begin{equation}
\begin{aligned}\begin{aligned}\ket[\psi_{j}]= & \sin\left(2j+1\right)\theta\left|t\right\rangle +\cos\left(2j+1\right)\theta\left|t_{\perp}\right\rangle \end{aligned}
.\end{aligned}
\end{equation}
So, it can be seen that the state of the quantum system always lies
in the two-dimensional subspace spanned by $|t\rangle$ and $|t_{\perp}\rangle$
during the repetitions of the Grover iteration, and the effect of
the algorithm is essentially to rotate the quantum system from the
initial state $|\psi\rangle$ towards the target state $t$.

\begin{figure}
\includegraphics[scale=0.16]{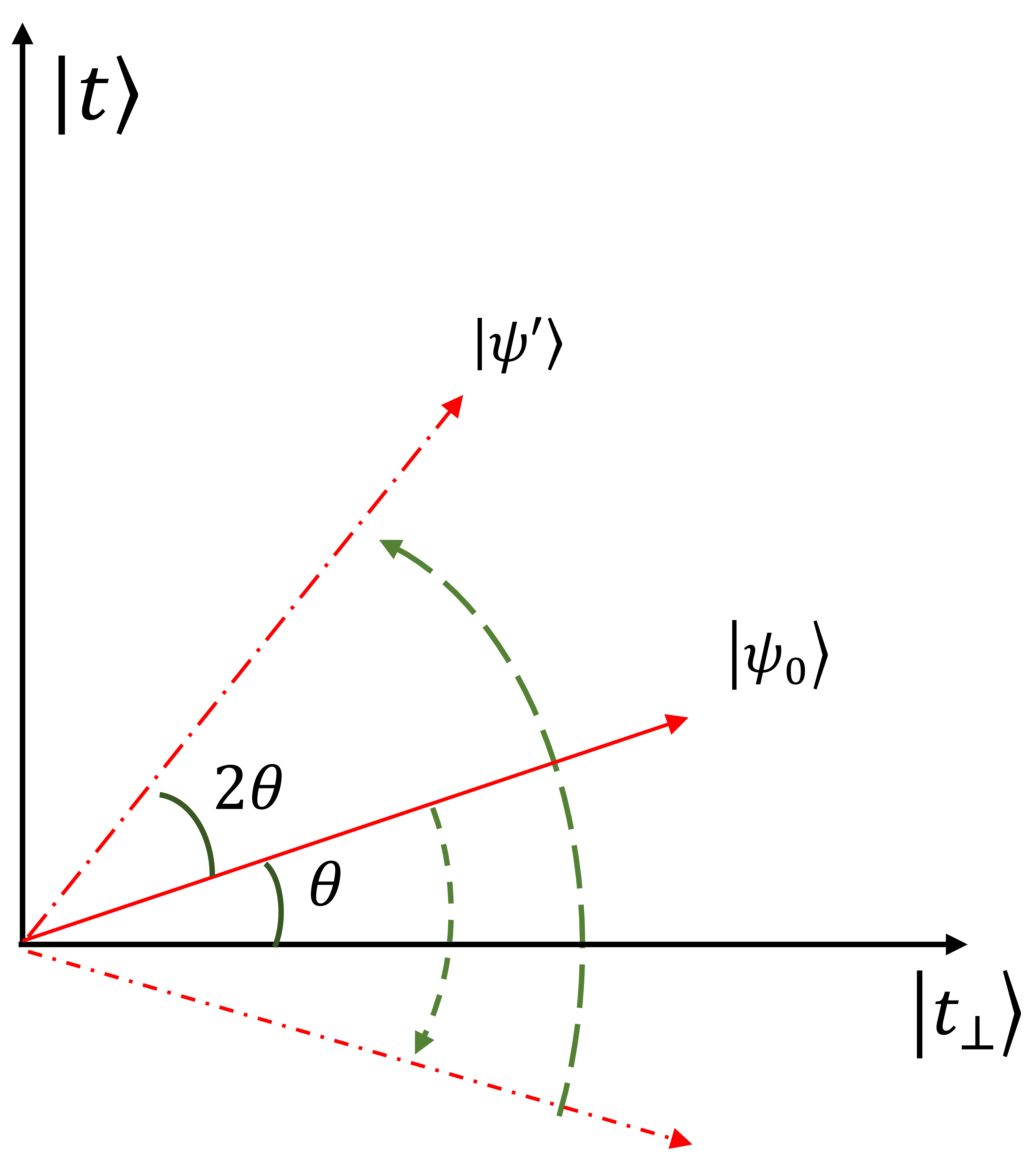}

\caption{Sketch of the standard Grover quantum search algorithm. The initial
state of the quantum system and the reflection axis of the diffusion
operator are both the uniform superposed state $\protect\ket[\psi_{0}]$
of all database elements. The initial state $\protect\ket[\psi_{0}]$
and the target state $\protect\ket[t]$ of the search problem span
a two-dimensional subspace. The Grover iteration consists of two steps:
First perform the oracle operation which reflects the system about
the state $\protect\ket[t_{\perp}]$ that is orthogonal to the target
state $\protect\ket[t]$ in the two-dimensional subspace, and then
perform the diffusion operation which reflects the system about the
state $\protect\ket[\psi_{0}]$. The total effect of the Grover iteration
is to rotate the system in the two-dimensional subspace away from
$\protect\ket[t_{\perp}]$ by double the angle between $\protect\ket[\psi_{0}]$
and $\protect\ket[t_{\perp}]$.}
\label{fig.pic1}
\end{figure}

The geometric picture of Grover's algorithm is illustrated in Fig.\,\ref{fig.pic1}.

\subsection{Query complexity of Grover's search algorithm}

After the evolution, one measures the quantum system along the computational
basis. If the system collapses to the target state, the search task
is completed successfully. In order to obtain the solution of the
search problem with a high probability, the quantum system should
be as close to the target state as possible at the end of the evolution.
Ideally, one expects to have the probability of obtaining the target
state
\begin{equation}
P_{j}=\left|\langle t\ket[\psi_{j}]\right|^{2}=\sin^{2}\left(2j+1\right)\theta=1,\label{eq:pj}
\end{equation}
so the optimal number of Grover iterations is
\begin{equation}
j=\frac{\pi}{4\arcsin\frac{1}{\sqrt{N}}}-\frac{1}{2}.
\end{equation}
When the size of the database, $N$, is large, $j$ can be approximated
as
\begin{equation}
j\approx\frac{\pi}{4}\sqrt{N}.
\end{equation}
In reality, as the number of Grover iterations needs to be an integer,
$j$ can usually be chosen as
\begin{equation}
j\approx\left\lceil \frac{\pi}{4}\sqrt{N}\right\rceil ,
\end{equation}
where $\left\lceil x\right\rceil $ is the ceiling function which
outputs the minimum integer that is no smaller than $x$.

\section{Optimization method\label{sec:Optimization method}}

When the elements of a database are equally likely to be the solution
of the search problem, Grover's quantum search algorithm has been
proven to be optimal in the query complexity. However, if the elements
have a nonuniform \emph{a priori} probability distribution to be the
search target, Grover's algorithm can be further improved, as one
may increase the weights of the basis states with higher probabilities
in the initial state of the quantum system so that the system can
approach the target state faster.

In this section, we study the minimization of the query complexity
of Grover's quantum search algorithm by optimizing the initial state
of the system and the reflection axis of the diffusion operator, provided
the average success probability of the algorithm to find the solution
is given.

\subsection{Success probability of generalized Grover search algorithm}

Consider a database of $N$ items, the \emph{a priori} probabilities
of which to be the search target is known. Denote the \emph{a priori}
probability of the $k$th element to be search target as $p_{k}$,
and the probabilities $p_{k}$, $k=1,\cdots,N$, are normalized,
\begin{equation}
p_{1}+\cdots+p_{N}=1.
\end{equation}
In contrast to the uniformly superposed initial state in the standard
Grover quantum search algorithm, a nonuniformly superposed initial
state of the quantum system may perform better when the \emph{a priori}
probabilities of the database elements are given, as one may increase
the weights of the basis states with higher \emph{a priori} probabilities
to accelerate the search algorithm. So, we assume the initial state
of the quantum system to be an arbitrary state in the current problem,
i.e.,
\begin{equation}
\left|\psi\right\rangle =\sum_{i=1}^{N}a_{i}\left|i\right\rangle ,
\end{equation}
where $a_{i}$'s are arbitrary coefficients that satisfy the normalization
condition,
\begin{equation}
|a_{1}|^{2}+\cdots|a_{N}|^{2}=1.
\end{equation}
Similarly, the reflection axis of the diffusion operator is not necessarily
the uniformly superposed state, as one may choose the reflection axis
to make the diffusion more beneficial to those basis states with higher
\emph{a priori} probabilities, so the reflection axis is also assumed
to be an arbitrary state in the current problem, i.e.,
\begin{equation}
\left|\varphi\right\rangle =\sum_{i=1}^{N}b_{i}\left|i\right\rangle ,
\end{equation}
where $b_{i}$'s are arbitrary coefficients satisfying the normalization
condition,
\begin{equation}
|b_{1}|^{2}+\cdots|b_{N}|^{2}=1.
\end{equation}

\begin{figure}
\includegraphics[scale=0.13]{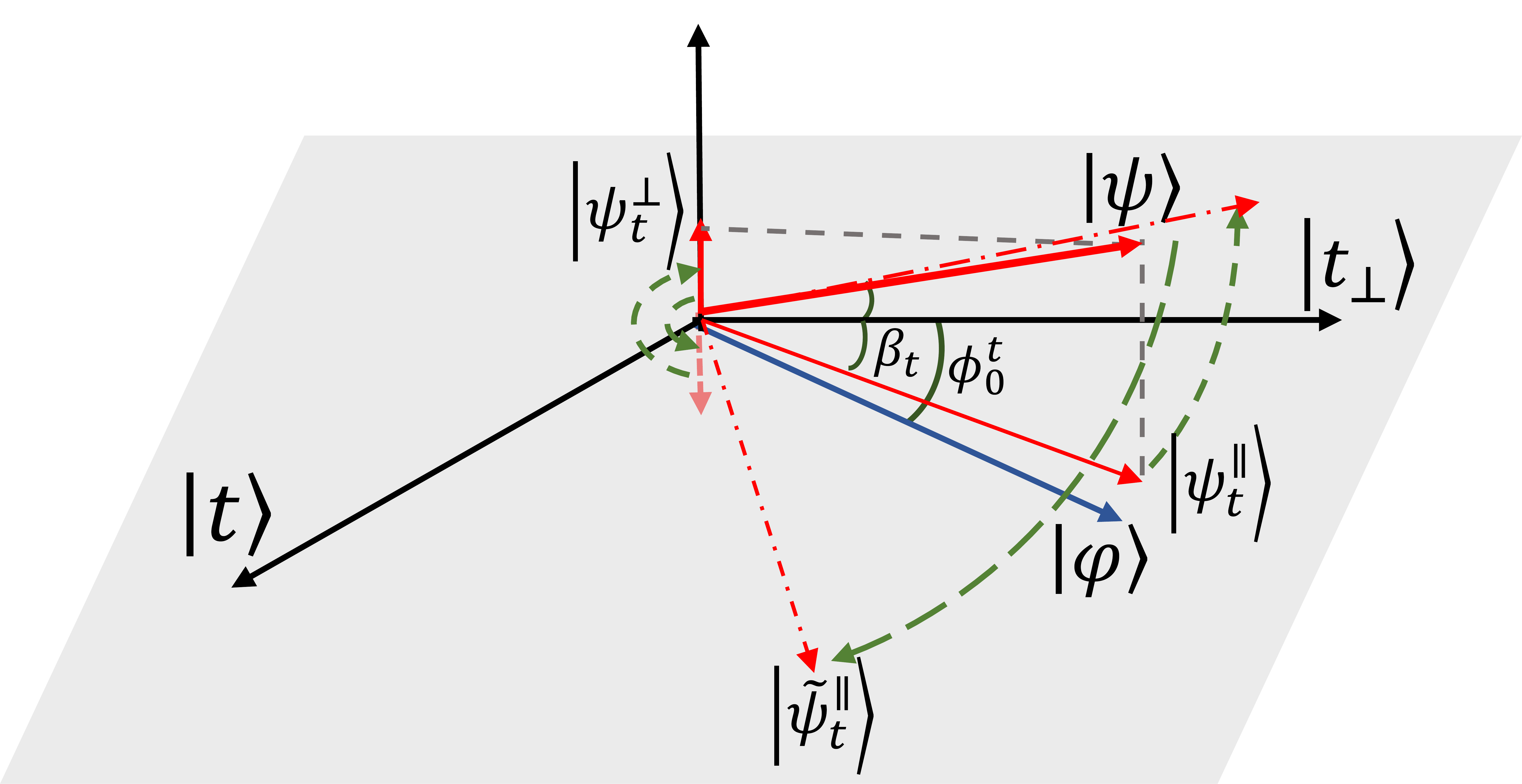}

\caption{Sketch of the improved quantum search algorithm. The initial state
$\protect\ket[\psi]$ of the quantum system and the reflection axis
$\protect\ket[\varphi]$ of the diffusion operator are assumed to
be arbitrary and do not necessarily coincide. The reflection axis
$\protect\ket[\varphi]$ and the solution state $\protect\ket[t]$
of the search problem span a two-dimensional subspace, and the state
of the quantum system can be decomposed into two components, $\vert\psi_{t}^{\parallel}\rangle$
parallel to this two-dimensional subspace and $\vert\psi_{t}^{\perp}\rangle$
orthogonal to it. The Grover iteration consists of two steps: First
perform the oracle operation which reflects the system about the state
$\protect\ket[t_{\perp}]$ that is orthogonal to the target state
$\protect\ket[t]$ in the two-dimensional subspace, and then perform
the diffusion operation which reflects the system about the reflection
axis $\protect\ket[\varphi]$. The total effect of the Grover iteration
is to rotate the parallel component $\vert\psi_{t}^{\parallel}\rangle$
of the system state in the two-dimensional subspace which is similar
to the original Grover search algorithm but with an additional flip
of the orthogonal component $\vert\psi_{t}^{\perp}\rangle$ about
the two-dimensional subspace.}
\label{fig.pic2}
\end{figure}

As the initial state of the system and the reflection axis of the
diffusion operator do not necessarily coincide, the state is no longer
rotating in a two-dimensional subspace during the Grover iterations
as in the standard Grover search algorithm. In contrast, the initial
state of the quantum system can now be decomposed into two orthogonal
components, one lying in the two-dimensional subspace spanned by the
target state and the reflection axis of the diffusion operator and
the other orthogonal to the two-dimensional subspace. The parallel
component (that lies within the two-dimensional subspace) is still
rotating in the two-dimensional subspace towards the target state
by Grover iterations, but the orthogonal component is just flipped
about the two-dimensional subspace by each Grover iteration and always
kept orthogonal to the two-dimensional subspace. So, we only need
to consider the parallel component of the system state in computing
the success probability of the algorithm in the following. The mechanism
of how the system state is changed by the generalized Grover iterations
is illustrated in Fig. \ref{fig.pic2}.

In order to obtain the parallel component of the system state, we
need to first find the projector onto the two-dimensional subspace
spanned by the target state $|t\rangle$ and the reflection axis $|\varphi\rangle$.
The projection operator can be derived by the Gram-Schmidt orthogonalization
of $|t\rangle$ and $|\varphi\rangle$, and the result turns out to
be
\begin{equation}
P_{t}=\frac{\lvert t\rangle\langle t\lvert+\lvert\varphi\rangle\langle\varphi\lvert-\lvert t\rangle\langle\varphi\lvert\langle t\lvert\varphi\rangle-\lvert\varphi\rangle\langle t\lvert\langle\varphi\lvert t\rangle}{1-\left|\langle t\lvert\varphi\rangle\right|^{2}},
\end{equation}
where the subscript $t$ of the projection operator $P_{t}$ denotes
that the projection operator depends on the target state $|t\rangle$.
With this projection operator, the parallel component of the system
initial state $|\psi\rangle$ can be obtained as
\begin{equation}
\vert\psi_{t}^{\parallel}\rangle=P_{t}\vert\psi\rangle=\sqrt{\langle\psi\vert P_{t}\vert\psi\rangle}(\sin\phi_{0}^{t}\left|t\right\rangle +\cos\phi_{0}^{t}\left|t^{\perp}\right\rangle ),
\end{equation}
where $\left|t^{\perp}\right\rangle $ is the state orthogonal to
the target state $|t\rangle$ in the two-dimensional subspace and
$\phi_{0}^{t}$ is the initial angle between the parallel component
and $\left|t^{\perp}\right\rangle $,
\begin{equation}
\phi_{0}^{t}=\arcsin\frac{\langle t\vert P_{t}\lvert\psi\rangle}{\sqrt{\langle\psi\vert P_{t}\vert\psi\rangle}}.
\end{equation}

Therefore, if the target state of the search problem is $|t\rangle$,
the success probability to find the system in the target state after
$j$ repetitions of the generalized Grover iteration is
\begin{equation}
P_{\mathrm{sus}}^{(t)}=\langle\psi\vert P_{t}\lvert\psi\rangle\sin^{2}\left(2j\beta_{t}+\phi_{0}^{t}\right),\label{eq:pt}
\end{equation}
where $\beta_{t}=\arcsin\langle t\lvert\varphi\rangle$ is the rotation
angle of the system state by a single Grover iteration, and both the
probability of projecting the system state onto the two-dimensional
subspace and the success probability of the final measurement to find
the target state are considered. If we also take the \emph{a priori}
probabilities of different database items into account, the final
success probability of the generalized Grover search algorithm turns
to be
\begin{equation}
\begin{aligned}\bar{P}= & \sum_{t}p_{t}P_{\mathrm{sus}}^{(t)}=\sum_{t}p_{t}\langle\psi\vert P_{t}\vert\psi\rangle\\
 & \times\sin^{2}\left(2j\arcsin\langle t\lvert\varphi\rangle+\arcsin\frac{\langle t\vert P_{t}\vert\psi\rangle}{\sqrt{\langle\psi\vert P_{t}\vert\psi\rangle}}\right).
\end{aligned}
\label{eq:Pbar}
\end{equation}
Eq. (\ref{eq:Pbar}) will be critical to the optimization of Grover's
search algorithm below.

\subsection{{\normalsize{}Optimization conditions}}

Now, we proceed to find the minimum number of oracle calls that can
drive the quantum system to the target state. It can be verified that
the standard Grover search algorithm is always optimal, whatever the
\emph{a priori} probabilities of the database items are, provided
the success probability of the search algorithm is required to be
$1$ (neglecting the integer nature of the number of the oracle calls).
Therefore, we allow a nonzero failure probability of the search algorithm
in this work, and investigate how the reduction in the number of oracle
calls can compensate for the loss in the success probability of the
search algorithm.

Suppose the success probability of the search algorithm is fixed as
$P_{0}$ and the number of oracle calls to realize this success probability
of the search algorithm is $j$. By the Lagrange multiplier method,
we can minimize the number of oracle calls $j$ by letting the variation
of the following function be zero,
\begin{equation}
F=j^{2}+\mu\left(\bar{P}-P_{0}\right)+\nu_{1}\left(\langle\psi\vert\psi\rangle-1\right)+\nu_{2}\left(\langle\varphi\vert\varphi\rangle-1\right),\label{eq:F}
\end{equation}
where $\mu$, $\nu_{1}$ and $\nu_{2}$ denote the Lagrange multipliers
for the constraint conditions of the success probability, the normalization
of the initial state and the normalization of the reflection axis
of the diffusion operator respectively. Note that $j$ must be positive,
so we minimize $j^{2}$ instead of $j$ in the above function (otherwise
the variation may generate a negative $j$ with the absolute value
minimized).

The number of oracle calls $j$ also needs to be varied in the variation
of $F$; however, the discreteness of $j$ makes the variation of
$j$ difficult. To circumvent this issue, we assume the number of
the database elements $N$ is large and renormalize the number of
oracle calls to
\begin{equation}
\lambda=2j\arcsin\frac{1}{\sqrt{N}},\label{eq:lam}
\end{equation}
which is still discrete in principle but can vary approximately in
a continuous way when $N$ is sufficiently large. The optimized $\lambda$
and the corresponding $j$ will generally be a float number, but one
just needs to take the ceiling function of $j$ to make $j$ an integer
which will change $j$ by no more than $1$, a negligible change when
$N$ is large, so we will just assume $\lambda$ to be a continuous
positive number in the following computation. The average success
probability $\bar{P}$ of the generalized Grover search algorithm
can be rewritten in terms of $\lambda$ as
\begin{equation}
\begin{aligned}\bar{P}= & \sum_{t}\left[p_{t}\langle\psi\vert P_{t}\vert\psi\rangle\right.\\
 & \left.\times\sin^{2}\left(\lambda\frac{\arcsin\langle t\lvert\varphi\rangle}{\arcsin\frac{1}{\sqrt{N}}}+\arcsin\frac{\langle t\vert P_{t}\vert\psi\rangle}{\sqrt{\langle\psi\vert P_{t}\vert\psi\rangle}}\right)\right],
\end{aligned}
\label{eq:P}
\end{equation}
and the $j^{2}$ term in $F$ (\ref{eq:F}) should be replaced by
$\lambda^{2}/(4\arcsin^{2}\frac{1}{\sqrt{N}})$ in this case.

The variation of $F$ includes the variation of the average success
probability as well as the other constraint conditions. Since the
average success probability is the main constraint condition in the
variation of $F$, we study the variation of the success probability
$\bar{P}$ first. By some computation, the variation of $\bar{P}$
can be written as
\begin{equation}
\delta\bar{P}=\langle\delta\psi\vert\text{\ensuremath{a_{\psi}\rangle}}+\langle\delta\varphi\vert\text{\ensuremath{b_{\varphi}\rangle}}+c_{\lambda}\delta\lambda.\label{eq:deltaP}
\end{equation}
As the expressions for $\vert a_{\psi}\rangle$, $\vert b_{\varphi}\rangle$
and $c_{\lambda}$ are quite lengthy, we leave the details to Appendix
\ref{sec:Differential-expansion}.

In Eq. (\ref{eq:deltaP}), there should have been Hermitian conjugate
terms of $\langle\delta\psi\vert\text{\ensuremath{a_{\psi}\rangle}}$
and $\langle\delta\varphi\vert\text{\ensuremath{b_{\varphi}\rangle}}$
in the variation of $\bar{P}$ (\ref{eq:deltaP}), but note that both
$\vert\psi\rangle$ and $|\varphi\rangle$ must be real states, so
there are only $\langle\delta\psi\vert a_{\psi}\rangle$ and $\langle\delta\varphi\vert b_{\varphi}\rangle$
in Eq. (\ref{eq:deltaP}). The reality of $\vert\psi\rangle$ and
$|\varphi\rangle$ is not a simplification here, but rather a property
of the current method: We start from the average success probability
of the quantum search algorithm and optimize the number of Grover
iterations by a variational approach, which will be further solved
by a differential method below that takes the standard Grover algorithm
as the initial parameters. Since the average success probability,
the variational approach and the differential method do not introduce
any imaginary coefficients and the standard Grover algorithm also
does not include any imaginary parameters, the resulted optimal initial
state \textit{$|\psi\rangle$ }and\textit{ }reflection axis \textit{$|\varphi\rangle$}
must always be real. Hence, the Hermitian conjugates of $\langle\delta\psi\vert a_{\psi}\rangle$
and $\langle\delta\varphi\vert b_{\varphi}\rangle$ coincide with
themselves.

Taking the other two constraint conditions as well as the $\lambda^{2}$
term into account, the full variation of $F$ can be obtained as
\begin{equation}
\begin{aligned}\delta F= & \langle\delta\psi\vert(2\nu_{1}\vert\psi\rangle+\mu\vert a_{\psi}\rangle)+\langle\delta\varphi\vert(2\nu_{2}\vert\varphi\rangle+\mu\vert b_{\varphi}\rangle)\\
 & +\left(\mu c_{\lambda}+\frac{\lambda}{2\arcsin^{2}\frac{1}{\sqrt{N}}}\right)\delta\lambda.
\end{aligned}
\end{equation}
When the number of oracle calls is minimized, the variation of $F$
should be zero, so this immediately leads to the following optimization
equations for the initial state $|\psi\rangle$, the reflection axis
$|\varphi\rangle$ of the diffusion operator, and the renormalized
number of oracle calls $\lambda$,
\begin{equation}
\begin{aligned}2\nu_{1}\vert\psi\rangle+\mu\vert a_{\psi}\rangle & =0,\\
2\nu_{2}\vert\varphi\rangle+\mu\vert b_{\varphi}\rangle & =0,\\
\mu c_{\lambda}+\frac{2\lambda}{4\arcsin^{2}\frac{1}{\sqrt{N}}} & =0.
\end{aligned}
\label{eq:variation eqs}
\end{equation}
Since $\mu$ can be chosen arbitrarily by changing $\nu_{1}$ and
$\nu_{2}$ accordingly in the first two equations of (\ref{eq:variation eqs}),
the third equation can then always be satisfied and does not need
to be further considered in the following computation.

By projecting the first two lines of (\ref{eq:variation eqs}) onto
the state $|\psi\rangle$ and $|\varphi\rangle$ respectively, one
can obtain the Lagrange multipliers $\nu_{1}$ and $\nu_{2}$,
\begin{equation}
\nu_{1}=-\frac{\mu}{2}\langle\psi\vert a_{\psi}\rangle,\;\nu_{2}=-\frac{\mu}{2}\langle\varphi\vert b_{\varphi}\rangle.
\end{equation}
Therefore, the optimization equations for $\vert a_{\psi}\rangle$
and $\vert b_{\varphi}\rangle$ can be finally obtained as
\begin{equation}
\begin{aligned}\vert a_{\psi}\rangle & =\langle\psi\vert a_{\psi}\rangle\vert\psi\rangle,\\
\vert b_{\varphi}\rangle & =\langle\varphi\vert b_{\varphi}\rangle\vert\varphi\rangle.
\end{aligned}
\label{eq:apbp}
\end{equation}
implying the proportionality between $\vert a_{\psi}\rangle$ and
$\vert\psi\rangle$ and between $\vert b_{\varphi}\rangle$ and $\vert\varphi\rangle$.

Eq. (\ref{eq:apbp}) is the optimization condition derived from the
Lagrange multipliers method for the initial state of the system, the
reflection axis of the diffusion operator and the renormalized number
of oracle calls. It will be the starting point of the study in the
following sections, from which one can obtain the minimal number of
oracle calls given the success probability of the search algorithm
and the corresponding optimized initial state and reflection axis.

\subsection{Differential solution to optimization equations\label{subsec:Differential-solution-to}}

Solving Eq. (\ref{eq:apbp}) is generally difficult, as the equation
is nonlinear with respect to the initial state, the reflection axis
and the renormalized number of oracle calls. In order to simplify
the problem, we assume the failure probability of the algorithm is
low, i.e., the success probability $\bar{P}$ is close to $1$, so
that the renormalized number of oracle calls and the initial state
and the reflection axis have only slight deviations from those of
the standard Grover search algorithm. In this case, we just need to
obtain the differential relation between the initial state, the reflection
axis and the renormalized number of oracle calls to minimize the query
complexity of the quantum search algorithm.

In the following, we give a formal differential solution to the optimization
problem based on this idea. In detail, one can take the differentiation
of Eq. (\ref{eq:apbp}), which produces
\begin{equation}
\begin{aligned}\vert da_{\psi}\rangle & =\langle d\psi\vert a_{\psi}\rangle\vert\psi\rangle+\langle\psi\vert da_{\psi}\rangle\vert\psi\rangle+\langle\psi\vert a_{\psi}\rangle\vert d\psi\rangle,\\
\vert db_{\varphi}\rangle & =\langle d\varphi\vert b_{\varphi}\rangle\vert\varphi\rangle+\langle\varphi\vert db_{\varphi}\rangle\vert\varphi\rangle+\langle\varphi\vert b_{\varphi}\rangle\vert d\varphi\rangle.
\end{aligned}
\end{equation}
Projecting these two equations onto $|\psi\rangle$ and $|\varphi\rangle$
respectively, one can obtain $\langle d\psi\vert a_{\psi}\rangle=0$
and $\langle d\varphi\vert b_{\varphi}\rangle=0$ by noting that $\langle\psi|d\psi\rangle=0$
and $\langle\varphi|d\varphi\rangle=0$ as $|\psi\rangle$ and $|\varphi\rangle$
are real and normalized states. Hence, the above two differential
equations can be simplified to
\begin{equation}
\begin{aligned}\left(I-\vert\psi\rangle\langle\psi\vert\right)\vert da_{\psi}\rangle & =\langle\psi\vert a_{\psi}\rangle\vert d\psi\rangle,\\
\left(I-\vert\varphi\rangle\langle\varphi\vert\right)\vert db_{\varphi}\rangle & =\langle\varphi\vert b_{\varphi}\rangle\vert d\varphi\rangle,
\end{aligned}
\label{eq:op}
\end{equation}
where $I$ denotes the $N\times N$ identity matrix.

The explicit results of $\vert a_{\psi}\rangle$ and $\vert b_{\varphi}\rangle$
can be derived from the variation of the average success probability
$\bar{P}$ as defined in Eq. (\ref{eq:deltaP}) and are shown in detail
in Appendix \ref{sec:Differential-expansion}, so their differentials
$\vert da_{\psi}\rangle$ and $\vert db_{\varphi}\rangle$ can be
obtained accordingly, which can be further expanded to the differentials
of $|\psi\rangle$, $|\varphi\rangle$ and $\lambda$,
\begin{equation}
\vert da_{\psi}\rangle=A_{\psi\psi}\vert d\psi\rangle+A_{\psi\varphi}\vert d\varphi\rangle+d\lambda\vert a_{\psi\lambda}\rangle,\label{eq:dap}
\end{equation}
where $A_{\psi\psi}$ and $A_{\psi\varphi}$ are both $N\times N$
matrices and $\vert a_{\psi\lambda}\rangle$ is an unnormalized $N\times1$
vector. Plugging Eq. (\ref{eq:dap}) into the first line of (\ref{eq:op}),
one can have
\begin{equation}
\begin{aligned}\left(I-\vert\psi\rangle\langle\psi\vert\right)\vert da_{\psi}\rangle= & \left(I-\vert\psi\rangle\langle\psi\vert\right)A_{\psi\psi}\vert d\psi\rangle\\
 & +\left(I-\vert\psi\rangle\langle\psi\vert\right)A_{\psi\varphi}\vert d\varphi\rangle\\
 & +\left(I-\vert\psi\rangle\langle\psi\vert\right)d\lambda\vert a_{\psi\lambda}\rangle\\
= & \langle\psi\vert a_{\psi}\rangle\vert d\psi\rangle.
\end{aligned}
\end{equation}
which can be rearranged to
\begin{equation}
\widetilde{A_{\psi}}\vert d\psi\rangle+\widetilde{A_{\varphi}}\vert d\varphi\rangle=d\lambda\vert v_{a}\rangle,
\end{equation}
where
\begin{equation}
\begin{aligned}\widetilde{A_{\psi}} & =\langle\psi\vert a_{\psi}\rangle I-\left(I-\vert\psi\rangle\langle\psi\vert\right)A_{\psi\psi},\\
\widetilde{A_{\varphi}} & =-\left(I-\vert\psi\rangle\langle\psi\vert\right)A_{\psi\varphi},\\
\vert v_{a}\rangle & =\left(I-\vert\psi\rangle\langle\psi\vert\right)\vert a_{\psi\lambda}\rangle.
\end{aligned}
\label{eq:at}
\end{equation}

Similarly, the differential $\vert db_{\varphi}\rangle$ can also
be expanded to the differentials of $|\psi\rangle$, $|\varphi\rangle$
and $\lambda$,
\begin{equation}
\vert db_{\varphi}\rangle=B_{\varphi\psi}\vert d\psi\rangle+B_{\varphi\varphi}\vert d\varphi\rangle+d\lambda\vert b_{\varphi\lambda}\rangle,\label{eq:dbp}
\end{equation}
where $B_{\varphi\psi}$ and $B_{\varphi\varphi}$ are $N\times N$
matrices and $\vert b_{\varphi\lambda}\rangle$ is an unnormalized
$N\times1$ vector. Plugging Eq. (\ref{eq:dbp}) into the second line
of (\ref{eq:op}) and rearranging the equation gives
\begin{equation}
\widetilde{B_{\varphi}}|d\varphi\rangle+\widetilde{B_{\psi}}\vert d\psi\rangle=d\lambda\vert v_{b}\rangle,
\end{equation}
where
\begin{equation}
\begin{aligned}\widetilde{B_{\psi}} & =-\left(I-\vert\varphi\rangle\langle\varphi\vert\right)B_{\varphi\psi},\\
\widetilde{B_{\varphi}} & =\langle\varphi\vert b_{\varphi}\rangle I-\left(I-\vert\varphi\rangle\langle\varphi\vert\right)B_{\varphi\varphi},\\
\vert v_{b}\rangle & =\left(I-\vert\varphi\rangle\langle\varphi\vert\right)\vert b_{\varphi\lambda}\rangle.
\end{aligned}
\label{eq:bt}
\end{equation}

Now, the two optimization conditions in Eq. (\ref{eq:op}) can be
merged and written in a more compact way,
\begin{equation}
\begin{aligned}M\begin{bmatrix}\vert d\psi\rangle\\
\vert d\varphi\rangle
\end{bmatrix}=\begin{bmatrix}\vert v_{a}\rangle\\
\vert v_{b}\rangle
\end{bmatrix}d\lambda\end{aligned}
,\label{eq:Mat}
\end{equation}
where $M$ is a $2N\times2N$ matrix,
\begin{equation}
M=\begin{bmatrix}\widetilde{A_{\psi}} & \widetilde{A_{\varphi}}\\
\widetilde{B_{\psi}} & \widetilde{B_{\varphi}}
\end{bmatrix}.\label{eq:M}
\end{equation}
Therefore, $\vert d\psi\rangle$ and $\vert d\varphi\rangle$ is given
by
\begin{equation}
\begin{bmatrix}\vert d\psi\rangle\\
\vert d\varphi\rangle
\end{bmatrix}=M^{-1}\begin{bmatrix}\vert v_{a}\rangle\\
\vert v_{b}\rangle
\end{bmatrix}d\lambda.\label{eq:dpsidphi}
\end{equation}
This is the formal differential relation between $|\psi\rangle$,
$|\varphi\rangle$ and $\lambda$.

With this differential relation, $\vert da_{\psi}\rangle$ (\ref{eq:dap})
and $\vert db_{\varphi}\rangle$ (\ref{eq:dbp}) can be written as
\begin{equation}
\begin{bmatrix}\vert da_{\psi}\rangle\\
\vert db_{\varphi}\rangle
\end{bmatrix}=\left(TM^{-1}\begin{bmatrix}\vert v_{a}\rangle\\
\vert v_{b}\rangle
\end{bmatrix}+\begin{bmatrix}\vert a_{\psi\lambda}\rangle\\
\vert b_{\varphi\lambda}\rangle
\end{bmatrix}\right)d\lambda\label{eq:dapsidbfi}
\end{equation}
where $T$ is a $2N\times2N$ matrix given by
\begin{equation}
T=\begin{bmatrix}A_{\psi\psi} & A_{\psi\varphi}\\
B_{\varphi\psi} & B_{\varphi\varphi}
\end{bmatrix}.\label{eq:T}
\end{equation}
Thus, the differential relations between $|a_{\psi}\rangle$, $|b_{\varphi}\rangle$
and $\lambda$ are also obtained.

As will be shown later, we will also need $\vert d^{2}\psi\rangle$
and $\vert d^{2}\varphi\rangle$ to compute the deviation of the success
probability of the search algorithm from the original Grover search
algorithm, so we obtain a formal solution to $\vert d^{2}\psi\rangle$
and $\vert d^{2}\varphi\rangle$ below. By taking differentiation
of $\vert d\psi\rangle$ and $\vert d\varphi\rangle$ in Eq. (\ref{eq:dpsidphi})
and noting that
\begin{equation}
\frac{dM^{-1}}{d\lambda}=-M^{-1}\frac{dM}{d\lambda}M^{-1},
\end{equation}
which can be derived from the differentiation of $M^{-1}M=I$, one
obtains
\begin{equation}
\begin{bmatrix}\vert d^{2}\psi\rangle\\
\vert d^{2}\varphi\rangle
\end{bmatrix}=M^{-1}\left(-\frac{dM}{d\lambda}M^{-1}\begin{bmatrix}\vert v_{a}\rangle\\
\vert v_{b}\rangle
\end{bmatrix}+\begin{bmatrix}\vert\frac{dv_{a}}{d\lambda}\rangle\\
\vert\frac{dv_{b}}{d\lambda}\rangle
\end{bmatrix}\right)d\lambda^{2}.
\end{equation}
This is the formal solution to $\vert d^{2}\psi\rangle$ and $\vert d^{2}\varphi\rangle$.
Note that $M$, $\vert v_{a}\rangle$ and $\vert v_{b}\rangle$ also
rely on $\ket[\psi]$ and $\ket[\varphi]$, so, Eq. (\ref{eq:dpsidphi})
will need to be invoked in the computation of $\frac{dM}{d\lambda}$,
$\vert\frac{dv_{a}}{d\lambda}\rangle$ and $\vert\frac{dv_{b}}{d\lambda}\rangle$.

Now, we can proceed to find the relation between the reduction of
the success probability of the quantum search algorithm and the decrease
in the number of oracle calls by expanding the average success probability
$\bar{P}$ to the differentials of $|\psi\rangle$, $|\varphi\rangle$
and $\lambda$. As the original Grover search algorithm gives $\bar{P}=1$
which is the maximal value of $\bar{P}$, the expansion of success
probability $\bar{P}$ to the first-order differentials of $|\psi\rangle$,
$|\varphi\rangle$ and $\lambda$ must be zero, so we need to consider
the second order differentials of $|\psi\rangle$, $|\varphi\rangle$
and $\lambda$. The differentiation of $\bar{P}$ is
\begin{equation}
d\bar{P}=\langle d\psi\vert\ensuremath{a_{\psi}\rangle}+\langle d\varphi\vert\ensuremath{b_{\varphi}\rangle}+c_{\lambda}d\lambda,\label{eq:dp}
\end{equation}
similar to the variation of $\bar{P}$ (\ref{eq:deltaP}), so, the
expansion of $\bar{P}$ to the second-order differentials $|\psi\rangle$,
$|\varphi\rangle$ and $\lambda$ can be obtained as
\begin{equation}
\begin{aligned}d\bar{P}= & \frac{1}{2}\big(\langle d\psi\vert da_{\psi}\rangle+\langle d\varphi\vert db_{\varphi}\rangle+\langle d^{2}\psi\vert a_{\psi}\text{\ensuremath{\rangle}}\\
 & +\langle d^{2}\varphi\vert b_{\varphi}\rangle+dc_{\lambda}d\lambda\big).
\end{aligned}
\label{eq:second}
\end{equation}

The differentials in Eq. (\ref{eq:second}) are carried out with $|\psi\rangle$,
$|\varphi\rangle$ and $\lambda$ given by the original Grover search
algorithm, so that the differentials of $|\psi\rangle$, $|\varphi\rangle$
and $\lambda$ represent small deviations from those in the original
Grover search algorithm. The matrices involved in the above computation,
e.g., $A_{\psi\psi}$, $A_{\psi\varphi}$, $B_{\varphi\psi}$, $B_{\varphi\varphi}$
and $\widetilde{A_{\psi}}$, $\widetilde{A_{\varphi}}$, $\widetilde{B_{\psi}},\widetilde{B_{\varphi}}$,
etc., are derived in Appendix \ref{sec:Derivation-of-the-1}. Using
the results in that appendix and plugging the differential relations
(\ref{eq:dpsidphi}) and (\ref{eq:dapsidbfi}) into Eq. (\ref{eq:second}),
one obtains
\begin{equation}
d\bar{P}=\frac{1}{2}\ord Sd\lambda^{2},\label{eq:dpdlam}
\end{equation}
where
\begin{equation}
\begin{aligned}\ord S= & \Big(-\bra[V]M^{\dagger-1}\frac{dM^{\dagger}}{d\lambda}M^{\dagger-1}\ket[\eta]+\bra[\frac{d}{d\lambda}V]M^{\dagger-1}\ket[\eta]\\
 & \ord{+\bra[V]M^{\dagger-1}TM^{-1}\ket[V]+\bra[V]M^{\dagger-1}\ket[\gamma]+\frac{dc_{\lambda}}{d\lambda}\Big)},
\end{aligned}
\label{eq:s}
\end{equation}
with $\ket[V]$, $\ket[\eta]$, $\ket[\gamma]$ short for
\begin{equation}
\ket[V]=\begin{bmatrix}\vert v_{a}\rangle\\
\vert v_{b}\rangle
\end{bmatrix},\;\ket[\eta]=\begin{bmatrix}\vert a_{\psi}\rangle\\
\vert b_{\varphi}\rangle
\end{bmatrix},\;\ket[\gamma]=\ket[\partial_{\lambda}\eta]=\begin{bmatrix}\vert a_{\psi\lambda}\rangle\\
\vert b_{\varphi\lambda}\rangle
\end{bmatrix}
\end{equation}
and the subscript ``std'' indicating the terms are evaluated with
$|\psi\rangle$, $|\varphi\rangle$ and $\lambda$ given by the standard
Grover search algorithm, i.e.,
\begin{equation}
\ord{|\psi\rangle}=\ord{|\varphi\rangle}=|\psi_{0}\rangle,\;\ord{\lambda}=\frac{\pi}{2}-\arccsc\sqrt{N},
\end{equation}
where $|\psi_{0}\rangle$ is the uniformly superposed state given
in (\ref{eq:psi0}).

Eq. (\ref{eq:dpdlam}) immediately gives the decrease of the renormalized
number of oracle calls in terms of the decrease in the success probability
of the quantum search algorithm when the latter is small,
\begin{equation}
\Delta\lambda=-\sqrt{\frac{2\Delta\bar{P}}{\ord S}}.\label{eq:dlp}
\end{equation}
The matrices and vectors in $\ord S$ (\ref{eq:s}) are obtained above
in Appendix \ref{sec:Derivation-of-the-1} with the parameters from
the standard Grover quantum search algorithm, except $M^{-1}$, $\frac{dM}{d\lambda}$
and $\vert dv_{a,b}\rangle$. Here $M^{-1}$ does not have a general
compact solution since $M$ is an $N\times N$ matrix, and $\frac{dM}{d\lambda}$
and $\vert dv_{a,b}\rangle$ rely on $M^{-1}$ since they involve
$\ket[d\psi]$ and $\ket[d\varphi]$ which needs $M^{-1}$ to be obtained
by (\ref{eq:dpsidphi}). But as $M$ is known (with the four block
submatrices given by (\ref{eq:abt})), $M^{-1}$ can be obtained once
$N$ is given, and thus $\frac{dM}{d\lambda}$ and $\vert dv_{a,b}\rangle$
can also be obtained accordingly.

Therefore, we finally arrive at a formal solution to the optimal number
of oracle calls given the decrease in the success probability of the
quantum search algorithm $\Delta P$,
\begin{equation}
j_{{\rm min}}=\left\lceil \frac{\arcsec\sqrt{N}-\sqrt{\frac{2\Delta P}{\ord S}}}{2\arccsc\sqrt{N}}\right\rceil ,\label{eq:jopt}
\end{equation}
where the renormalized number of oracle calls $\lambda$ has been
restored to the actual number of oracle calls $j$ by Eq. (\ref{eq:lam}).
When $N$ is large, $j_{\min}$ is approximately
\begin{equation}
j_{{\rm min}}\approx\left\lceil \sqrt{N}\left(\frac{\pi}{4}-\sqrt{\frac{\Delta P}{\ord{2S}}}\right)\right\rceil .
\end{equation}
The corresponding optimized initial state of the system and the reflection
axis of diffusion operator can be obtained from Eq. (\ref{eq:dpsidphi}),
\begin{equation}
\begin{bmatrix}|\psi_{\mathrm{opt}}\rangle\\
|\varphi_{\mathrm{opt}}\rangle
\end{bmatrix}=\begin{bmatrix}|\psi_{0}\rangle\\
|\psi_{0}\rangle
\end{bmatrix}+\ord{M^{-1}}\ord{\ket[V]}\sqrt{\frac{2\Delta P}{\ord S}}.\label{eq:psifhiopt}
\end{equation}

A subtle point in the above optimization approach for the quantum
search algorithm is that if the inversion of the matrix $M$ is carried
out numerically, the time complexity is usually $O(N^{\epsilon})$,
$\epsilon>2$, which is higher than the cost of the quantum search
algorithm and seems to eliminate the advantage of quantum search algorithm.
However, it should be noted that for an arbitrary database size $N$,
once $N$ is given, one can always work out the inversion of the matrix
$M$ analytically, as the matrix inversion just involves additions
and multiplications. Once the inversion of the matrix is derived,
it can be used repetitively in the above optimization of quantum search
algorithm, and the computation of matrix inversion does not need to
be invoked each time. So, the cost of the matrix inversion is just
one-time, and the advantage of the above optimized quantum search
algorithm with \emph{a priori} information does not vanish when the
algorithm runs for multiple times (e.g, for different choices of the
\emph{a priori} probability distribution).

\noindent \emph{Remark. }Eq. (\ref{eq:dpdlam}) implies that if $|\ord S|$
is small, a large increase in the number of oracle calls can only
increase a small portion of the success probability. So, if a low
failure probability of the quantum search algorithm is allowed, the
query complexity of the algorithm can be significantly decreased,
compared to the standard Grover quantum search algorithm. Certainly,
the failure probability of the quantum search algorithm will require
more trials of the algorithm to find the search target which may conversely
increase the query complexity of the algorithm. However, as the failure
probability is $\Delta P$ while the reduction in the number of oracle
calls is of order $O(\sqrt{\Delta P})$, the reduction in the oracle
calls is much more than the increase of oracle calls caused by the
failure probability, so the above optimization method can still lower
the query complexity of the quantum search algorithm on average.

\section{Example: two-value \emph{a priori} probability distribution\label{sec:Analysis and Results}}

In the preceding section, we derived the minimized number of oracle
calls for the quantum search algorithm, the optimized initial state
of the quantum system , and the optimized reflection axis of the diffusion
operator. In this section, we apply these results to a simple database
model to show how the \emph{a priori} knowledge of the search target
can assist in reducing the query complexity of quantum search algorithm.

Consider a database of $N$ elements. Suppose we know the \emph{a
priori} probabilities of the database elements to be the solution
of the search problem, and the \emph{a priori} probabilities are two
valued. For example, $K$ elements in the database have \emph{a priori}
probabilities $p$ to be the search solution and the other $N-K$
elements have \emph{a priori} probabilities
\begin{equation}
q=\frac{1-Kp}{N-K},
\end{equation}
to be the search solution. A low failure probability of the quantum
search algorithm $\Delta P$ is allowed.

The method introduced in the preceding section can be invoked to minimize
the number of oracle calls to reach the given success probability
of the search algorithm by optimizing the initial state of the quantum
system and the reflection axis of the diffusion operator.

\subsection{Approximate optimal solution}

The mathematical detail of the derivation is left to Appendix \ref{sec:Derivation-of-the}.
The factor $\ord S$ in the differential relation between the success
probability and the renormalized number of oracle calls (\ref{eq:dpdlam})
turns out to be
\begin{equation}
\ord S=\frac{2Np(Kp-1)}{-2Kp+K/N+Np}.\label{eq:S}
\end{equation}
Hence if the failure probability of the quantum search algorithm is
$\Delta P$, the decrease in the number of oracle calls is approximately
\begin{equation}
\Delta j=\frac{-1}{2\arccsc\sqrt{N}}\sqrt{\frac{\Delta P}{\ord S}}.\label{eq:dlam}
\end{equation}

\begin{figure}
\begin{centering}
\includegraphics[scale=0.45]{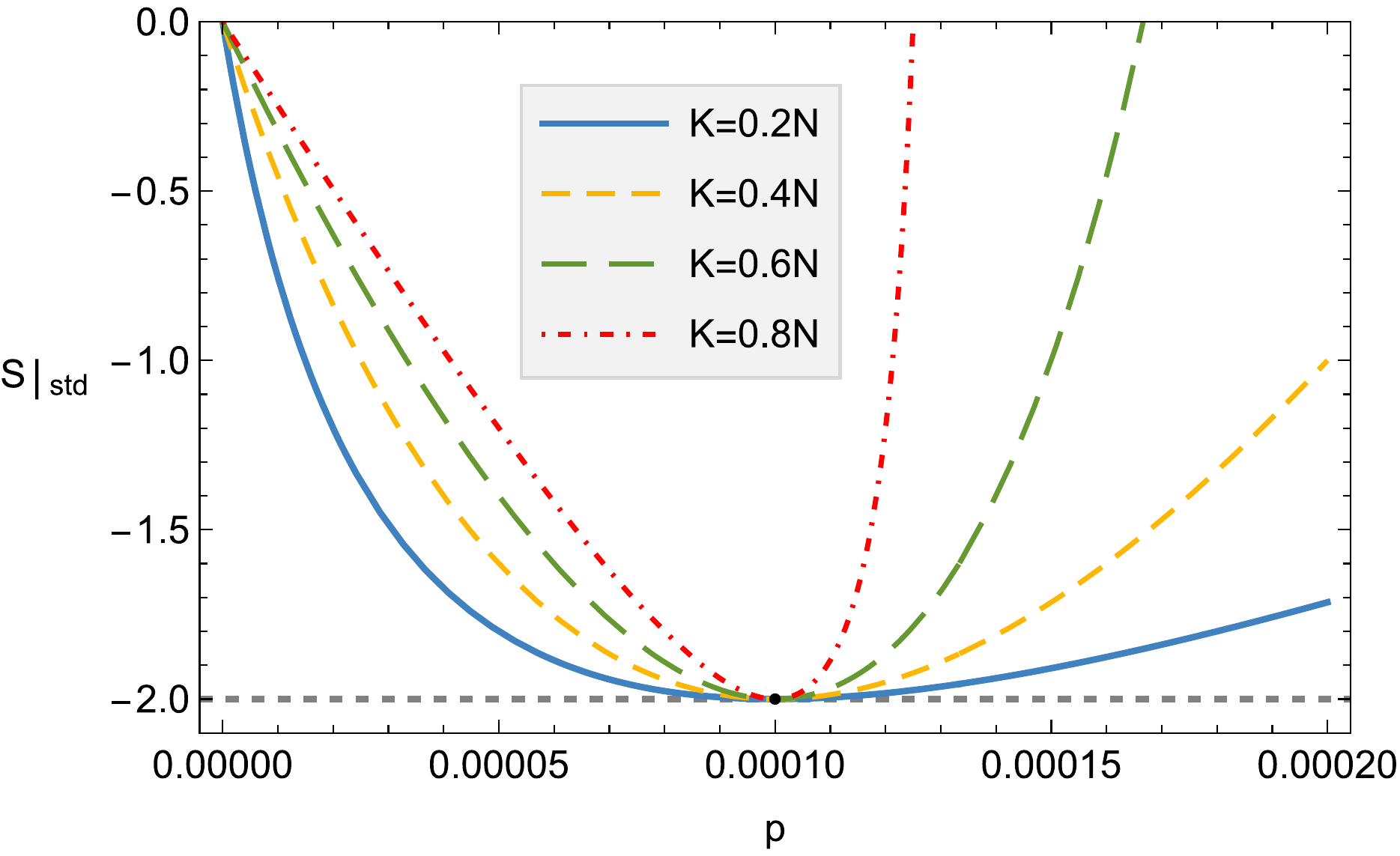}
\par\end{centering}
\caption{Plot of factor $\protect\ord S$ with respect to the \emph{a priori}
probability $p$ for the database model with two-value \emph{a priori}
probabilities. The number of database elements is $N=10^{4}$, and
the number of elements with \emph{a priori} probability $p$ is $K$.
The factor $\protect\ord S$ is plotted for different $K$. Note that
the range of $p$ varies with different $K$, as $Kp$ cannot exceed
the total \emph{a priori} probability $1$. It can be seen that all
the lines have the minimum value $-2$ at the point $p=10^{-4}$,
as this is the case of uniform \emph{a priori} probabilities corresponding
to the original Grover quantum search algorithm which is already optimal
and cannot be optimized anymore. All the other points with $p\protect\neq10^{-4}$
represent the cases of nonuniform \emph{a priori} probability distributions
which can be optimized by the approach presented in this paper, and
the smaller the factor $\protect\ord S$ is, the more the number of
Grover iterations can be reduced.}

\label{fig.con}
\end{figure}

To illustrate this result, the factor $S$ is plotted for different
values of $p$ and $K/N$ in Fig. \ref{fig.con}.

An interesting special case is that the \emph{a priori} probabilities
for different elements of the database are uniform, i.e.,
\begin{equation}
p=\frac{1}{N},
\end{equation}
which is exactly the case considered by the standard Grover search
algorithm. It can be obtained from Eq. (\ref{eq:S}) as well as observed
from Fig. \ref{fig.con} that for this case,
\begin{equation}
\ord S=-2,\label{eq:s-2}
\end{equation}
implying that the query complexity of the original Grover search algorithm
can also be decreased if a failure probability of the search algorithm
is allowed.

The above result $\ord S=-2$ for the original Grover search algorithm
can also be obtained in a straightforward way as follows, without
invoking the method introduced in the previous section. As the success
probability of the original Grover algorithm to find the solution
of the search problem with $j$ steps of Grover iterations is known
to be
\begin{equation}
P_{G}=\sin^{2}\left[\left(2j+1\right)\arccsc\sqrt{N}\right]=\sin^{2}\left(\lambda+\arccsc\sqrt{N}\right),\label{eq:pg}
\end{equation}
where $P_{G}$ denotes the success probability for the original Grover
search algorithm and $\lambda$ is the renormalized number of oracle
calls defined in (\ref{eq:lam}), one can obtain
\begin{equation}
\begin{aligned}\frac{dP_{G}}{d\lambda} & =\sin\left(2\lambda+2\arccsc\sqrt{N}\right),\\
\frac{d^{2}P_{G}}{d\lambda^{2}} & =2\cos\left(2\lambda+2\arccsc\sqrt{N}\right).
\end{aligned}
\end{equation}
For the original Grover search algorithm,
\begin{equation}
P_{G}=1,\;\lambda=\frac{\pi}{2}-\arccsc\sqrt{N},
\end{equation}
so, one can obtain that
\begin{equation}
\frac{dP_{G}}{d\lambda}=0,
\end{equation}
which can be understood from that $P_{G}$ is at the maximal value
$1$, and
\begin{equation}
\frac{d^{2}P_{G}}{d\lambda^{2}}=-2,
\end{equation}
which is in accordance with the result $\ord S=-2$ in Eq. (\ref{eq:s-2}).

This result for the original Grover search algorithm seems quite natural,
as a failure probability of the search algorithm can certainly allow
a reduction in the number of oracle calls. Also, note that the initial
state of the system and the reflection axis of the diffusion operator
are not optimized to obtain $\ord S=-2$ in this case, since it can
be verified that for the original Grover search algorithm, the uniformly
superposed state (\ref{eq:psi0}) is already the optimal choice for
the initial state and the reflection axis even when a failure probability
of the search algorithm is allowed.

However, this is not the case when the \emph{a priori} probabilities
of the database elements are not uniform. For nonuniform \emph{a priori}
probabilities, one needs to change the initial state of the system
and the reflection axis of the diffusion operator to minimize the
query complexity of the quantum search algorithm, and this is the
goal of the optimization method proposed in the previous section.
In fact, by the optimization of the initial state and the reflection
axis, the quantum search algorithm can gain more increase in the efficiency
for database elements with nonuniform \emph{a priori} probabilities
than with uniform \emph{a priori} probabilities. This can be observed
from Fig. (\ref{fig.con}) as $\ord S=-2$ of the original Grover
search algorithm has the largest absolute value over all possible
values of $\text{\ensuremath{\ord S}}$ for various \emph{a priori}
probabilities $p$.

As Eq. (\ref{eq:dlam}) indicates that a smaller $S$ results in a
larger decrease in the number of oracle calls, Fig. \ref{fig.con}
and Eq. (\ref{eq:dlam}) imply that all nonuniform \emph{a priori}
probability distributions of the database elements have a bigger boost
in the efficiency of the quantum search algorithm than the uniform
\emph{a priori} probability distribution for the original Grover quantum
search algorithm. This shows the advantage of a nonuniform \emph{a
priori} distribution of database elements in optimizing the performance
of quantum search algorithm, which is in accordance with the intuition
that one may exploit nonuniform \emph{a priori} probability distribution
of database elements to adjust the initial state of the system and
the reflection axis of the diffusion operator to be more beneficial
to the states with higher \emph{a priori} probabilities so that the
quantum search algorithm can have lower query complexity on average.

As remarked in the preceding section, though the optimization of the
quantum search algorithm relies on the decrease of the success probability
which may require more trials of the algorithm to find the target,
such an optimization approach can still increase the efficiency of
the search algorithm on average, as the decrease of the success probability
is of order $O(\Delta P)$ while the decrease of the number of oracle
calls is of order $O(\sqrt{\Delta P})$, the latter of which is much
larger when $\Delta P$ is small. The above result for the original
quantum search algorithm with uniform \emph{a priori} probabilities
can serve as an intuitive example of this point. It can be seen from
Eq. (\ref{eq:pg}) that when the quantum search algorithm is close
to completion, i.e., $\left(2j+1\right)\arcsin\frac{1}{\sqrt{n}}$
is close to $\frac{\pi}{2}$, the increase of the number of oracle
calls $j$ can hardly increase the success probability, so if a negligible
portion of success probability is dropped, a significant portion of
oracle iterations may be reduced! This is why the query complexity
of the original Grover search algorithm can be improved by the optimization
method.

\subsection{Discussion\label{subsec:Effectiveness-of-approximate}}

In the preceding subsection, we expanded the change of average success
probability $d\bar{P}$ of the quantum search algorithm to the second
order of $d\lambda$ which generates an approximate optimal solution
to the number of iterations in terms of $d\bar{P}$.. As we expand
$d\bar{P}$ to the lowest nonzero order, i.e., the second order, of
$d\lambda$, it is intuitive that if $d\bar{P}$ becomes large, this
approximation will break. So there exists an effective range of $d\bar{P}$
to validate the above approximate optimal solution.

In this subsection, we investigate the effectiveness of the above
approximate optimal solution by using the first-order optimal initial
state and reflection axis and expanding $d\bar{P}$ to the third-order
of $d\lambda$ to generate a more accurate optimal solution of the
renormalized iteration number. The deviation of this more accurate
optimal solution from that in the last subsection will give the effective
range of $d\bar{P}$.

As the approximate optimal solution is obtained in the vicinity of
$\bar{P}=1$ which is the maximal value of $\bar{P}$, the first-order
term in the expansion of $d\bar{P}$ must be zero, so the expansion
always starts from the second order of $d\lambda$. We plug the approximate
optimal solution to the initial state $|\psi\rangle$ and the reflection
axis $|\varphi\rangle$ of diffusion (\ref{eq:first-order-optimal})
(in terms of $\vert d\psi\rangle$ and $\vert d\varphi\rangle$) into
the average success probability $\bar{P}$ (\ref{eq:P}), and expand
the change of the success probability to the third order of $d\lambda$,
\begin{equation}
d\bar{P}=\frac{1}{2}Sd\lambda^{2}+\frac{1}{6}Gd\lambda^{3}.\label{eq:dpbar}
\end{equation}
From this expansion, one can work out $d\lambda$ in terms of $d\bar{P}$,
\begin{equation}
d\lambda=-\sqrt{\frac{2d\bar{P}}{\ord S}}+\ord Bd\bar{P},\label{eq:dl3}
\end{equation}
with $\ord S$ obtained in Eq. (\ref{eq:S}) and
\begin{align}
\ord B=-\frac{(K-1)K(N-2)(Np-1)^{2}\arccsc\sqrt{N}}{\pi\sqrt{N-1}Np(-2KNp+K+Np)(KNp-1)} & .
\end{align}
The mathematical details of the derivation can be found in Appendix
\ref{sec:Calculation of higher-order expansions}.

Eq. (\ref{eq:dl3}) is an extension of the result in the last subsection
to involve the first order of $d\bar{P}$. To validate the approximation,
this additional term should be much smaller than the term of order
$\frac{1}{2}$, so it leads to the effective range of $d\bar{P}$
for the approximation,
\begin{equation}
d\bar{P}\ll\frac{2}{\ord S(\ord B)^{2}}.
\end{equation}

What if $S$ is too small, say $S\ll B$? For example, one can see
from Fig. \ref{fig.con} that $S$ approaches zero when the \emph{a
priori} probability of some database elements is zero, i.e., $p=0$.
This seems peculiar at first glance, but is actually in accordance
with physics. Note that the portion of database elements with \emph{a
priori} probability $p=0$ can never be a solution of the quantum
search task, so the quantum search algorithm can be restricted to
the remaining part of the database which includes $N-K$ elements
in this case, which can immediately reduce the query complexity of
the quantum search algorithm from $O(\sqrt{N})$ to $O(\sqrt{N-K})$,
even without a drop of the success probability! This means that we
can have a nonzero $d\lambda$ when $d\bar{P}=0$ for this case, which
leads to $S=d\bar{P}/d\lambda^{2}=0$.

However, it should also be noted that when $S$ is too small, the
second-order approximation for the expansion of $d\bar{P}$ is no
longer sufficient, and one must keep higher-order terms in Eq. (\ref{eq:dpbar})
to derive $d\lambda$. The solution to $d\lambda$ for this case can
be much more complex than that with the second-order approximation,
but is still tractable as Eq. (\ref{eq:dpbar}) is always a polynomial
equation in essence.

\begin{figure}
\begin{centering}
\subfloat[]{\centering{}\includegraphics[scale=0.45]{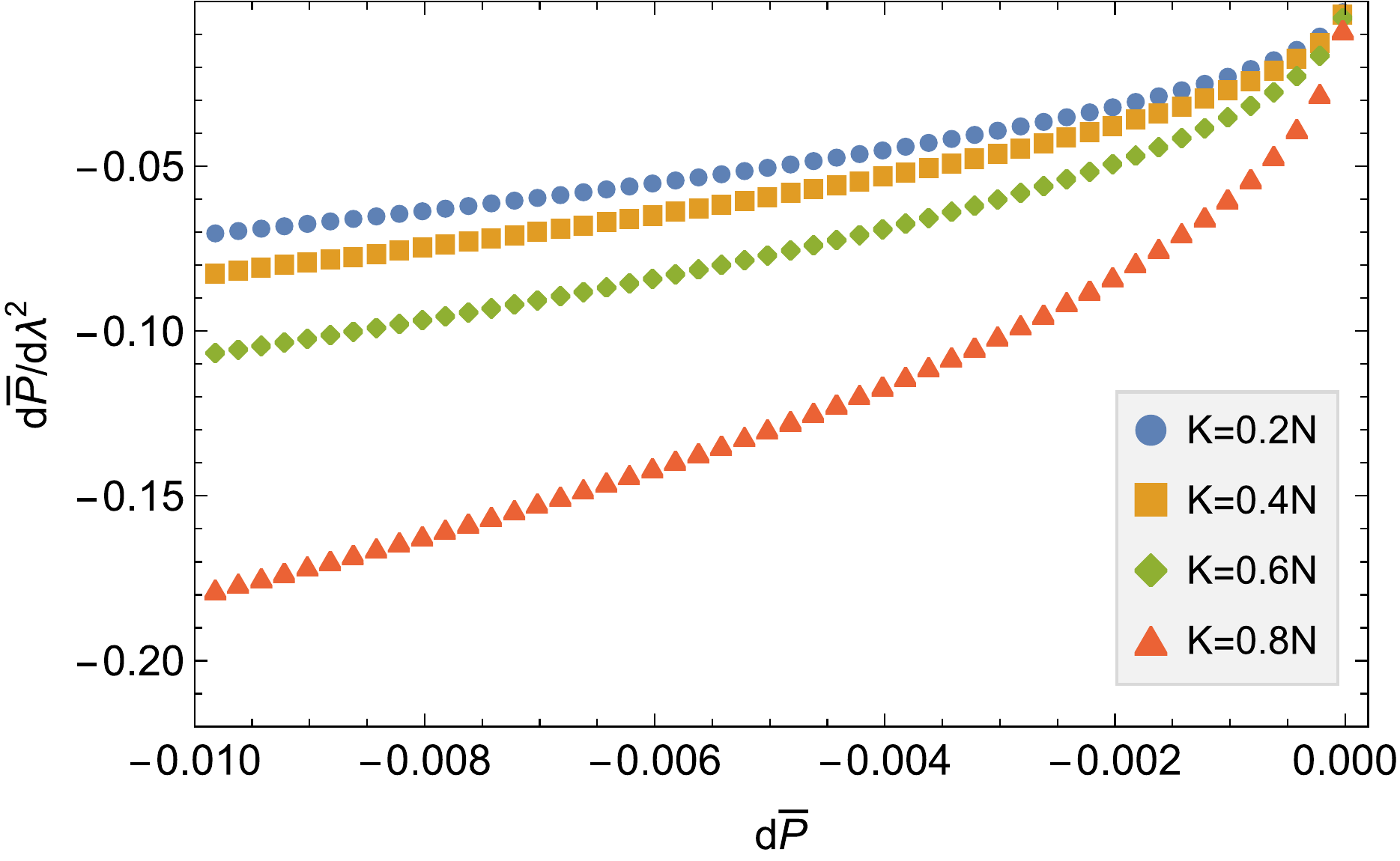}}
\par\end{centering}
\begin{centering}
\subfloat[]{\begin{centering}
\includegraphics[scale=0.45]{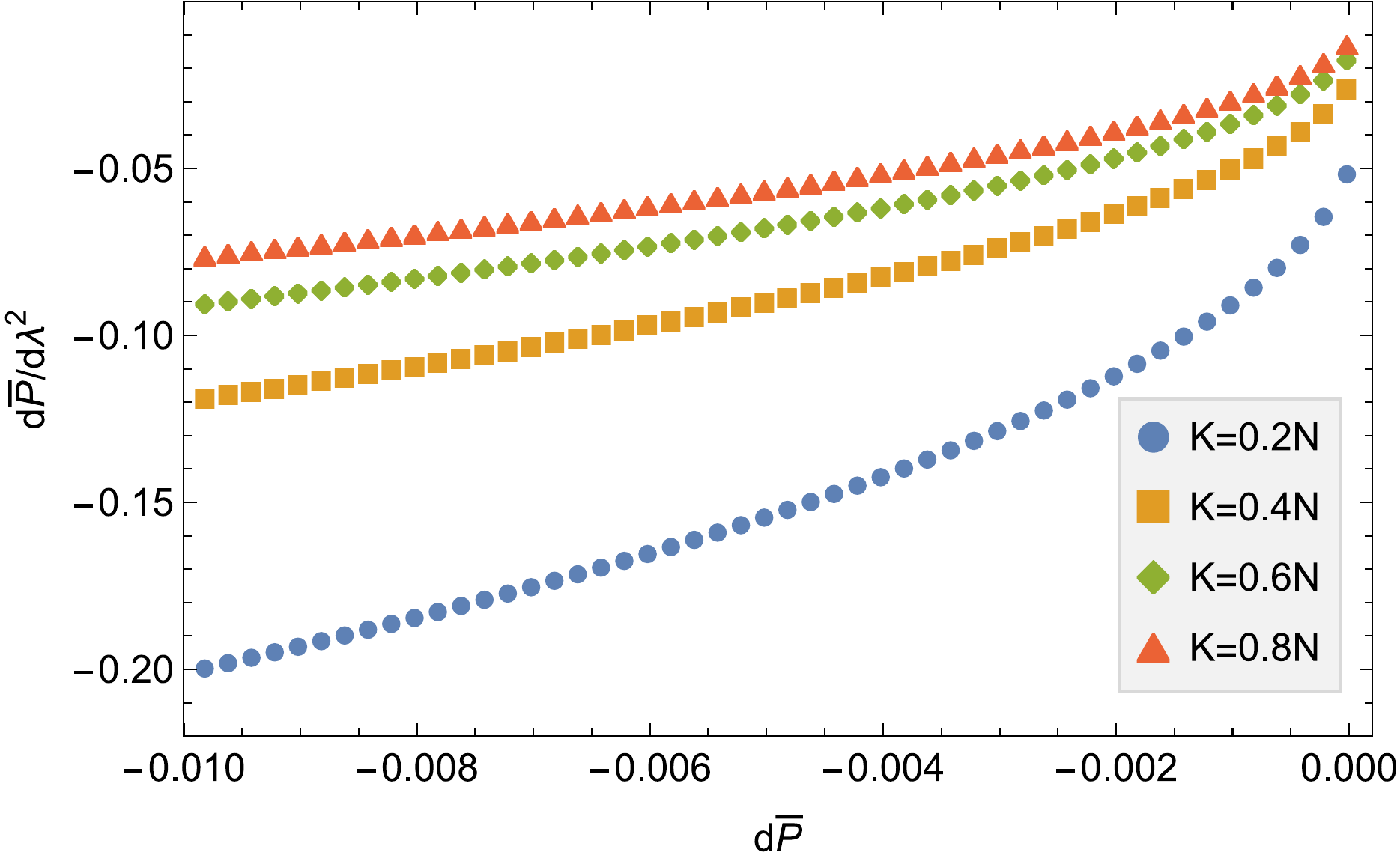}
\par\end{centering}
}
\par\end{centering}
\begin{centering}
\caption{Plots of the exact ratio of the change $d\bar{P}$ in the average
success probability to the square of the change $d\lambda$ in the
renormalized iteration steps for a database with different two-value
\emph{a priori} probability distributions: (a) $p=\left(1-10^{-6}\right)/K$
and (b) $p=10^{-6}$. The number of database elements is $N=10^{4}$,
and the number of elements with \emph{a priori} probability $p$ is
$K$. Lines with different colors indicate different choices of $K$.
The plots show how $d\bar{P}/d\lambda^{2}$ deviates from $\protect\ord S$
when $|d\bar{P}|$ becomes large and thus the necessity to keep higher-order
terms of $d\lambda$ in the expansion of $d\bar{P}$ for this scenario
as shown in Eq. (\ref{eq:dpbar}).}
\par\end{centering}
\label{fig.con2}
\end{figure}

To illustrate the effect of higher order terms on the approximation,
the exact ratio of the change $d\bar{P}$ in the average success probability
to the square of the change $d\lambda$ in the renormalized number
of iteration steps is plotted in Fig. \ref{fig.con2} for different
$p$ and $K$, which shows the deviation of $d\bar{P}/d\lambda^{2}$
from $\ord S$ when $|d\bar{P}|$ becomes large and thus the necessity
to include higher-order terms of $d\lambda$ in the expansion of $d\bar{P}$
for this case as shown in Eq. (\ref{eq:dpbar}).

\section{Conclusions\label{sec:Conclusion-and-outlook}}

In this work, we studied the quantum search algorithm with general
\emph{a priori} probabilities for the elements of the database to
be the solution of the search problem, and optimized the initial state
of the quantum system and the reflection axis of the diffusion operator
by a variational approach to minimize the number of oracle calls at
the expense of a failure probability of the quantum search algorithm.
{} We obtained the optimization conditions for the initial state and
the reflection axis, but the optimization equations are difficult
to solve due to being highly nonlinear, so we assumed the failure
probability of the quantum search algorithm to be low so that the
optimized initial state and reflection axis have just slight deviations
from those in the standard Grover search algorithm and thus a differentiation
approach can be exploited to obtain the differential relation between
the initial state of the quantum system, the reflection axis of the
diffusion operator, the renormalized number of Grover iterations and
the success probability of the search algorithm. We obtained a formal
approximate solution to the renormalized number of oracle calls in
terms of the decrease in the success probability, and applied it to
a simple database model with two-value \emph{a priori} probabilities
of the elements to exemplify the power of the optimization results.
The results showed that the optimization can indeed lower the query
complexity of the quantum search algorithm, even for the original
Grover search algorithm with uniform \emph{a priori} probabilities
for the database elements, which is not surprising as the success
probability of the search algorithm is lowered, but the optimization
can lead to a larger reduction in the query complexity when the \emph{a
priori} probability distribution is nonuniform, which suggests the
advantage of nonuniform \emph{a priori} probability distributions
in improving the efficiency of the quantum search algorithm.

We remark that the current results are effective for a small failure
probability, as we expanded the success probability only to the lowest-order
nonzero differential of the renormalized number of oracle calls. If
one wants to obtain results effective for larger failure probabilities,
higher-order differentials of the renormalized number of oracle calls
need to be included. The way to expand the success probability to
higher-order differentials of the renormalized number of oracle calls
is similar to the differentiation approach presented in this work.
The main difference lies in that the differential equation (\ref{eq:dpdlam})
will include multiple higher-order terms of the renormalized number
of oracle calls, so one will need to solve a polynomial equation to
obtain the solution for the decrease of the renormalized number of
oracle calls, which will increase the complexity of this optimization
problem if one wants to find analytical results but can be solved
efficiently by numerical computation.

We hope this work can provide a new perspective on the quantum search
problem, and stimulate further research in improving the efficiency
of quantum search algorithm by \emph{a priori} information.
\begin{acknowledgments}
The authors acknowledge helpful discussions with Jingyi Fan and Junyan
Li. This work is supported by the National Natural Science Foundation
of China (Grant No. 12075323).
\end{acknowledgments}

\appendix
\onecolumngrid

\section{Variation of success probability for generalized quantum search algorithm\label{sec:Differential-expansion}}

In combination with the prior distribution, the final success probability
$\bar{P}$ can be written as
\begin{equation}
\bar{P}=\sum_{t}p_{t}\langle\psi\vert P_{t}\vert\psi\rangle\sin^{2}\left(\lambda\frac{\arcsin\langle t\lvert\varphi\rangle}{\arccsc\sqrt{N}}+\arcsin\frac{\langle t\vert P_{t}\vert\psi\rangle}{\sqrt{\langle\psi\vert P_{t}\vert\psi\rangle}}\right)\label{eq:pb}
\end{equation}
where $\lambda=2j\arccsc\sqrt{N}$.

The variation of $\bar{P}$ can be written as
\begin{equation}
\delta\bar{P}=\langle\delta\psi\vert a_{\psi}\rangle+\langle\delta\varphi\vert b_{\varphi}\rangle+c_{\lambda}\delta\lambda.
\end{equation}
The unnormalized states $\vert a_{\psi}\rangle$ can be obtained by
taking the variation of $\bar{P}$ (\ref{eq:pb}) with respect to
$|\psi\rangle$, which turns out to be
\begin{equation}
\begin{aligned}\vert\text{\ensuremath{a_{\psi}\rangle=}} & \sum_{t}p_{t}\frac{\sqrt{1-b_{t}^{2}}(2a_{t}b_{t}-c)\sin\frac{2\lambda\arcsin b_{t}}{\arccsc\sqrt{N}}+\left[a_{t}\left(2b_{t}^{2}-1\right)-\text{\ensuremath{b_{t}}}c\right]\cos\frac{2\lambda\arcsin\text{\ensuremath{b_{t}}}}{\arccsc\sqrt{N}}-\text{\ensuremath{a_{t}}}+\text{\ensuremath{b_{t}}}c}{b_{t}^{2}-1}\lvert t\rangle\\
 & +\sum_{t}p_{t}\frac{-a_{t}\sqrt{1-\text{\ensuremath{b_{t}}}^{2}}\sin\frac{2\lambda\arcsin\text{\ensuremath{b_{t}}}}{\arccsc\sqrt{N}}+(c-\text{\ensuremath{a_{t}}}b_{t})\cos\frac{2\lambda\arcsin b_{t}}{\arccsc\sqrt{N}}+a_{t}\text{\ensuremath{b_{t}}}-c}{\text{\ensuremath{b_{t}}}^{2}-1}\lvert\varphi\rangle,
\end{aligned}
\label{eq:apsi}
\end{equation}
where
\begin{equation}
a_{t}=\langle t\vert\psi\rangle,\;b_{t}=\langle t\vert\varphi\rangle,\;c=\langle\varphi\vert\psi\rangle.
\end{equation}
Similarly, the unnormalized state $\vert\text{\ensuremath{b_{\varphi}\rangle}}$
can be obtained as
\begin{equation}
\begin{aligned}\vert\text{\ensuremath{b_{\varphi}\rangle=}} & \sum_{t}\frac{p_{t}}{\left(b_{t}^{2}-1\right)^{2}}\left\{ \frac{\lambda\left\{ \sqrt{1-\text{\ensuremath{b_{t}}}^{2}}\left[\text{\ensuremath{a_{t}}}^{2}\left(2\text{\ensuremath{b_{t}}}^{2}-1\right)-2\text{\ensuremath{a_{t}}}\text{\ensuremath{b_{t}}}c+c^{2}\right]\sin\frac{2\lambda\arcsin\text{\ensuremath{b_{t}}}}{\csc^{-1}\sqrt{n}}+2\text{\ensuremath{a_{t}}}\left(b_{t}^{2}-1\right)(a_{t}\text{\ensuremath{b_{t}}}-c)\cos\frac{2\lambda\arcsin b_{t}}{\arccsc\sqrt{N}}\right\} }{\arccsc\sqrt{N}}\right.\\
 & \left.-(a_{t}-\text{\ensuremath{b_{t}}}c)\left[a_{t}\sqrt{1-b_{t}^{2}}\sin\frac{2\lambda\arcsin b_{t}}{\arccsc\sqrt{N}}+(\text{\ensuremath{a_{t}}}\text{\ensuremath{b_{t}}}-c)\cos\frac{2\lambda\arcsin b_{t}}{\arccsc\sqrt{N}}-a_{t}b_{t}+c\right]\right\} \lvert t\rangle\\
 & +\sum_{t}p_{t}\frac{\text{\ensuremath{a_{t}}}\sqrt{1-\text{\ensuremath{b_{t}}}^{2}}\sin\frac{2\lambda\arcsin b_{t}}{\arccsc\sqrt{N}}+(\text{\ensuremath{a_{t}}}b_{t}-c)\cos\frac{2\lambda\arcsin b_{t}}{\arccsc\sqrt{N}}-\text{\ensuremath{a_{t}}}b_{t}+c}{1-b_{t}^{2}}\lvert\psi\rangle,
\end{aligned}
\label{bphi}
\end{equation}
and the coefficient $c_{\lambda}$ can be obtained as
\begin{equation}
c_{\lambda}=\sum_{t}p_{t}\arcsin\text{\ensuremath{b_{t}}}\frac{\left[\text{\text{\ensuremath{a_{t}}}}^{2}\left(1-2\text{\ensuremath{b_{t}}}^{2}\right)+2\text{\ensuremath{a_{t}}}\text{\ensuremath{b_{t}}}c-c^{2}\right]\sin\frac{2\lambda\arcsin\text{\ensuremath{b_{t}}}}{\arccsc\sqrt{N}}+2\text{\ensuremath{a_{t}}}\sqrt{1-\text{\ensuremath{b_{t}}}^{2}}(\text{\text{\ensuremath{a_{t}}}}\text{\ensuremath{b_{t}}}-c)\cos\frac{2\lambda\arcsin b_{t}}{\arccsc\sqrt{N}}}{\left(\text{\text{\ensuremath{b_{t}}}}^{2}-1\right)\arccsc\sqrt{N}}
\end{equation}

The differentials of $\vert a_{\psi}\rangle$, $\vert\text{\ensuremath{b_{\varphi}\rangle}}$
in Eq. (\ref{eq:dap}), (\ref{eq:dbp}) can be straightforwardly obtained
by taking differentiation of (\ref{eq:apsi}) and (\ref{bphi}), which
are not presented here as they are quite lengthy.

\section{Derivation of submatrices in $M$ and $T$ with parameters of original
Grover algorithm\label{sec:Derivation-of-the-1}}

In this appendix, we derive the matrices and vectors involved in the
computation in Sec. \ref{subsec:Differential-solution-to}, particularly
the submatrices of $M$ (\ref{eq:M}) and $T$ (\ref{eq:T}), i.e.,
$A_{\psi\psi}$, $A_{\psi\varphi}$, $B_{\varphi\psi}$, $B_{\varphi\varphi}$
and $\widetilde{A_{\psi}}$, $\widetilde{A_{\varphi}}$, $\widetilde{B_{\psi}},\widetilde{B_{\varphi}}$,
with the parameters of the standard Grover search algorithm, so that
the differentials of $|\psi\rangle$, $|\varphi\rangle$ and $\lambda$
represent the deviation from those in the standard Grover search algorithm.

By invoking the standard Grover quantum search algorithm and using
the results of $\vert a_{\psi}\rangle$, $|b_{\varphi}\rangle$ and
$\lambda$ in Appendix \ref{sec:Differential-expansion}, it can be
obtained that
\begin{equation}
\ord{\vert a_{\psi}\rangle}=\ord{\vert b_{\varphi}\rangle}=2\vert\psi_{0}\rangle,\;\ord{c_{\lambda}}=0,
\end{equation}
where the subscript ``std'' indicates the terms are evaluated with
$|\psi\rangle$, $|\varphi\rangle$ and $\lambda$ given by the standard
Grover search algorithm, i.e.,
\begin{equation}
\ord{|\psi\rangle}=\ord{|\varphi\rangle}=|\psi_{0}\rangle,\;\ord{\lambda}=\frac{\pi}{2}-\arccsc\sqrt{N},
\end{equation}
where $|\psi_{0}\rangle$ is the uniformly superposed state defined
in Eq. (\ref{eq:psi0}).

Taking the differentiation of $\vert a_{\psi}\rangle$, $|b_{\varphi}\rangle$,
$\lambda$ in Appendix \ref{sec:Differential-expansion} and evaluating
the resulting differentials with $|\psi\rangle$, $|\varphi\rangle$
and $\lambda$ from the standard Grover search algorithm, one obtains
\begin{equation}
\begin{aligned}\vert da_{\psi}\rangle & =\ord{A_{\psi\psi}}\vert d\psi\rangle+\ord{A_{\psi\varphi}}\vert d\varphi\rangle+\ord{\vert a_{\psi\lambda}\rangle}d\lambda,\\
\vert db_{\varphi}\rangle & =\ord{B_{\varphi\psi}}\vert d\psi\rangle+\ord{B_{\varphi\varphi}}\vert d\varphi\rangle+\ord{\vert b_{\varphi\lambda}\rangle}d\lambda,
\end{aligned}
\end{equation}
where
\begin{equation}
\begin{aligned}\ord{A_{\psi\psi}} & =2\pros[\psi_{0}]{\psi_{0}},\\
\ord{A_{\psi\varphi}} & =2\left(I+\pros[\psi_{0}]{\psi_{0}}\right)+\frac{\pi\sqrt{N}\pros[\psi_{0}]{\eta}-\pi N\sum_{t}p_{t}\pros[t]t}{(N-1)\arccsc\sqrt{N}},\\
\ord{\vert a_{\psi\lambda}\rangle} & =\frac{2\ket[\psi_{0}]-2\sqrt{N}\ket[\eta]}{\sqrt{N-1}},
\end{aligned}
\label{eq:aa}
\end{equation}
\begin{equation}
\begin{aligned}\ord{B_{\varphi\psi}} & =2\left(I+\pros[\psi_{0}]{\psi_{0}}\right)+\frac{\pi\sqrt{N}\pros[\eta]{\psi_{0}}-\pi N\sum_{t}p_{t}\pros[t]t}{(N-1)\arccsc\sqrt{N}},\\
\ord{B_{\varphi\varphi}} & =\frac{\pi N\left(\pi-4\arcsec\sqrt{N}\right)\sum_{t}p_{t}\pros[t]t}{2(N-1)\arccsc^{2}\sqrt{N}}+\frac{\pi\sqrt{N}\left(\pros[\psi_{0}]{\eta}+\pros[\eta]{\psi_{0}}\right)}{(N-1)\arccsc\sqrt{n}}+2\pros[\psi_{0}]{\psi_{0}},\\
\ord{\vert b_{\varphi\lambda}\rangle} & =\frac{2\ket[\psi_{0}]+2\sqrt{N}\ket[\eta]}{\sqrt{N-1}}-\frac{\sqrt{N}\pi\ket[\eta]}{\sqrt{N-1}\arccsc\sqrt{N}},
\end{aligned}
\label{eq:bb}
\end{equation}
and
\begin{equation}
\ket[\eta]=\sum_{t}p_{t}\ket[t],
\end{equation}
which is an unnormalized state determined by the \emph{a priori} probability
distribution $p_{t}$ of the database elements. With the above $\ord{A_{\psi\psi}}$,
$\ord{A_{\psi\varphi}}$, $\ord{B_{\varphi\psi}}$, $\ord{B_{\varphi\varphi}}$,
one can obtain the matrix $T$ for $\vert da_{\psi}\rangle$ and $\vert db_{\varphi}\rangle$
(\ref{eq:dapsidbfi}).

The matrices $\widetilde{A_{\psi}}$, $\widetilde{A_{\varphi}}$,
$\widetilde{B_{\psi}}$, $\widetilde{B_{\varphi}}$ can be also be
obtained accordingly by using the definitions (\ref{eq:at}) and (\ref{eq:bt})
as
\begin{equation}
\begin{aligned}\ord{\widetilde{A_{\psi}}} & =2I,\\
\ord{\widetilde{A_{\varphi}}} & =2\left(\pros[\psi_{0}]{\psi_{0}}-I\right)+\frac{\pi N\sum_{t}p_{t}\pros[t]t-\pi\sqrt{N}\pros[\psi_{0}]{\eta}}{(N-1)\arccsc\sqrt{N}},\\
\ord{\widetilde{B_{\psi}}} & =-2I+\frac{\pi N\sum_{t}p_{t}\pros[t]t-\pi\sqrt{N}\left(\pros[\psi_{0}]{\eta}+\pros[\eta]{\psi_{0}}\right)+\pi+2\pi(N-1)\arccsc\sqrt{N}\pros[\psi_{0}]{\psi_{0}}}{(N-1)\arccsc\sqrt{N}},\\
\ord{\widetilde{B_{\varphi}}} & =2I+\frac{\pi\left(\pros[\psi_{0}]{\psi_{0}}-\sqrt{N}\pros[\eta]{\psi_{0}}\right)}{(N-1)\arccsc\sqrt{N}}+\frac{\pi\left(\pi-4\arccsc\sqrt{N}\right)}{2(N-1)\arccsc^{2}\sqrt{N}}\left(N\sum_{t}p_{t}\pros[t]t-\sqrt{N}\pros[\psi_{0}]{\eta}\right).
\end{aligned}
\label{eq:abt}
\end{equation}
With $\ord{\widetilde{A_{\psi}}}$, $\ord{\widetilde{A_{\varphi}}}$,
$\ord{\widetilde{B_{\psi}}}$ and $\ord{\widetilde{B_{\varphi}}}$,
one can obtain the matrix $M$ for $\vert d\psi\rangle$ and $\vert d\varphi\rangle$
(\ref{eq:dpsidphi}).

In addition, the differential of $c_{\lambda}$ can also be obtained
with the parameters from the standard Grover quantum search algorithm,
\begin{equation}
\ord{dc_{\lambda}}=-2\left(\frac{\sqrt{N}\arcsec\sqrt{N}\langle\eta\vert\ensuremath{d\varphi\rangle}}{\sqrt{N-1}\arccsc\sqrt{N}}+\frac{\sqrt{N}\langle\eta\vert\ensuremath{d\psi\rangle}}{\sqrt{N-1}}+d\ensuremath{\lambda}\right),
\end{equation}
and the vectors $\vert v_{a}\rangle$, $\vert v_{b}\rangle$ in Eqs.
(\ref{eq:at}) and (\ref{eq:bt}) can be obtained as
\begin{equation}
\ord{\ket[v_{a}]}=\frac{2\ket[\psi_{0}]-2\sqrt{N}\ket[\eta]}{\sqrt{N-1}}d\lambda,\;\ord{\ket[v_{b}]}=\frac{\arcsec\sqrt{N}}{\arccsc\sqrt{N}}\ket[v_{a}].\label{eq:vv}
\end{equation}

\section{Derivation of the results for database with two-value \emph{a priori}
probabilities\label{sec:Derivation-of-the}}

In this appendix, we provide the mathematical details of the derivation
of the results for the database with two-value \emph{a priori} probabilities
in Sec. \ref{sec:Analysis and Results}.

Consider a database of $N$ elements, the \emph{a priori} probabilities
of which to be the search target are two valued. For example, there
are $K$ elements in the database with \emph{a priori} probabilities
$p$ to be the search target and the other $N-K$ elements with \emph{a
priori} probabilities $q$ to be the search target,
\begin{equation}
q=\frac{1-Kp}{N-K}.
\end{equation}

We use the method introduced in Appendix \ref{sec:Derivation-of-the-1}
to optimize the quantum search algorithm for this database model given
a small failure probability. The variation of the average success
probability with respect to the initial state $\ket[\psi]$, the reflection
axis $\ket[\varphi]$ of the diffusion operator and the renormalized
number $\lambda$ of oracle calls produces $|a_{\psi}\rangle$, $|b_{\varphi}\rangle$,
and $c_{\lambda}$ respectively, according to Eq. (\ref{eq:deltaP}),
\begin{align}
\ord{\vert a_{\psi}\rangle} & =\frac{2}{\sqrt{N}}E_{N\times1},\\
\ord{\vert b_{\varphi}\rangle} & =\frac{2}{\sqrt{N}}E_{N\times1},\\
\ord{c_{\lambda}} & =0,
\end{align}
with $\ket[\psi]$, $\ket[\varphi]$ and $\lambda$ given by the standard
Grover search algorithm, and $E_{n\times m}$ an $n\times m$ matrix
defined as
\begin{equation}
E_{n\times m}=\underset{m\,\mathrm{columns}}{\left.\underbrace{\begin{bmatrix}1 & 1 & \cdots & 1\\
1 & 1 & \cdots & 1\\
\vdots & \vdots & \ddots & \vdots\\
1 & 1 & \cdots & 1
\end{bmatrix}}\right\} }n\,\mathrm{rows}
\end{equation}
So, $E_{N\times1}$ is just an $N$-dimensional vector with all the
elements being 1's. For the sake of simplicity, we will also denote
$E_{k\times k}$ by $E_{k}$ in the following.

Taking the differentiation of $|a_{\psi}\rangle$, $|b_{\varphi}\rangle$,
and $c_{\lambda}$ with the parameters from the standard Grover search
algorithm, one can work out that
\begin{equation}
\begin{aligned}\ord{\vert da_{\psi}\rangle} & =\ord{A_{\psi\psi}}\vert d\varphi\rangle+\ord{A_{\psi\varphi}}\vert d\varphi\rangle+\ord{\vert a_{\psi\lambda}\rangle}d\lambda,\\
\ord{\vert db_{\varphi}\rangle} & =\ord{B_{\varphi\psi}}\ket[d\psi]+\ord{B_{\varphi\varphi}}\vert d\varphi\rangle+\ord{\vert b_{\varphi\lambda}\rangle}d\lambda,
\end{aligned}
\end{equation}
where
\begin{equation}
\begin{aligned}\ord{A_{\psi\psi}} & =\frac{2}{N}E_{N},\\
\ord{A_{\psi\varphi}} & =\begin{bmatrix}2(I_{K}+E_{K})+\frac{\pi p}{(N-1)\arccsc\sqrt{N}}\left(-NI_{K}+E_{K}\right) & \left[2+\frac{\pi q}{(N-1)\arccsc\sqrt{N}}\right]E_{K\times(N-K)}\\
\left[2+\frac{\pi p}{(N-1)\arccsc\sqrt{N}}\right]E_{(N-K)\times K} & 2(I_{N-K}+E_{N-K})+\frac{\pi q}{(N-1)\arccsc\sqrt{N}}\left(-NI_{N-K}+E_{N-K}\right)
\end{bmatrix},\\
\ord{\vert a_{\psi\lambda}\rangle} & =\begin{bmatrix}\left(\frac{2}{\sqrt{N\left(N-1\right)}}-\frac{2\sqrt{N}p}{\sqrt{N-1}}\right)E_{K\times1}\\
\left(\frac{2}{\sqrt{N\left(N-1\right)}}-\frac{2\sqrt{N}q}{\sqrt{N-1}}\right)E_{(N-K)\times1}
\end{bmatrix},
\end{aligned}
\end{equation}
and
\begin{equation}
\begin{aligned}\ord{B_{\varphi\psi}} & =\begin{bmatrix}2(I_{K}+E_{K})+\frac{\pi p}{(N-1)\arccsc\sqrt{N}}\left(-NI_{K}+E_{K}\right) & \left[2+\frac{\pi p}{(N-1)\arccsc\sqrt{N}}\right]E_{K\times(N-K)}\\
\left[2+\frac{\pi q}{(N-1)\arccsc\sqrt{N}}\right]E_{(N-K)\times K} & 2(I_{N-K}+E_{N-K})+\frac{\pi q}{(N-1)\arccsc\sqrt{N}}\left(-NI_{N-K}+E_{N-K}\right)
\end{bmatrix},\\
\ord{B_{\varphi\varphi}} & =\begin{bmatrix}\frac{\pi p\left(\pi-4\arccsc\sqrt{N}\right)N}{2(N-1)\arccsc^{2}\sqrt{N}}I_{K}+\left[2+\frac{2\pi p}{(N-1)\arccsc\sqrt{N}}\right]E_{K} & \left[2+\frac{\pi(p+q)}{(N-1)\arccsc\sqrt{N}}\right]E_{K\times(N-K)}\\
\left[2+\frac{\pi(p+q)}{(N-1)\arccsc\sqrt{N}}\right]E_{(N-K)\times K} & \frac{\pi q\left(\pi-4\arccsc\sqrt{N}\right)N}{2(N-1)\arccsc^{2}\sqrt{N}}I_{N-K}+\left[2+\frac{2\pi q}{(N-1)\arccsc\sqrt{N}}\right]E_{N-K}
\end{bmatrix},\\
\ord{\vert b_{\varphi\lambda}\rangle} & =\begin{bmatrix}\left(\frac{2}{\sqrt{N\left(N-1\right)}}+\frac{2\sqrt{N}p}{\sqrt{N-1}}-\frac{\sqrt{N}\pi p}{\sqrt{N-1}\arccsc\sqrt{N}}\right)E_{K\times1}\\
\left(\frac{2}{\sqrt{N\left(N-1\right)}}+\frac{2\sqrt{N}q}{\sqrt{N-1}}-\frac{\sqrt{N}\pi q}{\sqrt{N-1}\arccsc\sqrt{N}}\right)E_{(N-K)\times1}
\end{bmatrix}.
\end{aligned}
\end{equation}
With $\ord{A_{\psi\psi}}$, $\ord{A_{\psi\varphi}}$, $\ord{B_{\varphi\psi}}$,
$\ord{B_{\varphi\varphi}}$, one can obtain the matrix $T$ that is
necessary for the derivation of $\vert da_{\psi}\rangle$ and $\vert db_{\varphi}\rangle$
in (\ref{eq:dapsidbfi}).

The four block submatrices $\widetilde{A_{\psi}}$, $\widetilde{A_{\varphi}}$,
$\widetilde{B_{\psi}}$, $\widetilde{B_{\varphi}}$ for the matrix
$M$ in (\ref{eq:M}) can be also obtained with the parameters from
the standard quantum search algorithm as
\begin{equation}
\begin{aligned}\ord{\widetilde{A_{\psi}}} & =2I_{N},\\
\ord{\widetilde{A_{\varphi}}} & =\begin{bmatrix}\left[\frac{2}{N}-\frac{\pi p}{(N-1)\arccsc\sqrt{N}}\right]\left(E_{K}-NI_{K}\right) & \left[\frac{2}{N}-\frac{\pi q}{(N-1)\arccsc\sqrt{N}}\right]E_{K\times(N-K)}\\
\left[\frac{2}{N}-\frac{\pi p}{(N-1)\arccsc\sqrt{N}}\right]E_{(N-K)\times K} & \left[\frac{2}{N}-\frac{\pi q}{(N-1)\arccsc\sqrt{N}}\right]\left(E_{N-K}-NI_{N-K}\right)
\end{bmatrix},\\
\ord{\widetilde{B_{\psi}}} & =\begin{bmatrix}-2I_{K}+\frac{\pi+2\pi(N-1)\arccsc\sqrt{N}E_{K}}{N(N-1)\arccsc\sqrt{N}}+\frac{\pi p\left(NI_{K}-2E_{K}\right)}{(N-1)\arccsc\sqrt{N}} & \left[\frac{\pi+2\pi(N-1)\arccsc\sqrt{N}}{N(N-1)\arccsc\sqrt{N}}-\frac{\pi(p+q)}{(N-1)\arccsc\sqrt{N}}\right]E_{K\times(N-K)}\\
\left[\frac{\pi+2\pi(N-1)\arccsc\sqrt{N}}{N(N-1)\arccsc\sqrt{N}}-\frac{\pi(p+q)}{(N-1)\arccsc\sqrt{N}}\right]E_{(N-K)\times K} & -2I_{N-K}+\frac{\pi+2\pi(N-1)\arccsc\sqrt{N}E_{N-K}}{N(N-1)\arccsc\sqrt{N}}+\frac{\pi q\left(NI_{N-K}-E_{N-K}\right)}{(N-1)\arccsc\sqrt{N}}
\end{bmatrix}\\
\ord{\widetilde{B_{\varphi}}} & =\begin{bmatrix}2I_{K}+\frac{\pi\left(\frac{1}{N}-p\right)E_{K}}{(N-1)\arccsc\sqrt{N}}+\frac{\pi p\left(\pi-4\arccsc\sqrt{N}\right)\left(NI_{K}-E_{K}\right)}{2(N-1)\arccsc^{2}\sqrt{N}} & \left[\frac{\pi\left(\frac{1}{N}-p\right)}{(N-1)\arccsc\sqrt{N}}-\frac{\pi q\left(\pi-4\arccsc\sqrt{N}\right)}{2(N-1)\arccsc^{2}\sqrt{N}}\right]E_{K\times(N-K)}\\
\left[\frac{\pi\left(\frac{1}{N}-q\right)}{(N-1)\arccsc\sqrt{N}}-\frac{\pi p\left(\pi-4\arccsc\sqrt{N}\right)}{2(N-1)\arccsc^{2}\sqrt{N}}\right]E_{(N-K)\times K} & 2I_{N-K}+\frac{\pi\left(\frac{1}{N}-q\right)E_{N-K}}{(N-1)\arccsc\sqrt{N}}+\frac{\pi q\left(\pi-4\arccsc\sqrt{N}\right)\left(NI_{N-K}-E_{N-K}\right)}{2(N-1)\arccsc^{2}\sqrt{N}}
\end{bmatrix}
\end{aligned}
\end{equation}
With $\ord{\widetilde{A_{\psi}}}$, $\ord{\widetilde{A_{\varphi}}}$,
$\ord{\widetilde{B_{\psi}}}$ and $\ord{\widetilde{B_{\varphi}}}$,
one can obtain the matrix $M$ and its inverse that is necessary for
the derivation of $\vert d\psi\rangle$ and $\vert d\varphi\rangle$
in (\ref{eq:dpsidphi}).

According to the above results, the inverse of the matrix $M$ can
be found as
\begin{equation}
M^{-1}=\begin{bmatrix}\alpha_{1}(p)I_{K}+\beta_{1}(p,q)E_{K} & \gamma_{1}E_{K\times(N-K)} & \alpha_{2}(p)I_{K}+\beta_{2}(p,q)E_{K} & \gamma_{2}(p,q)E_{K\times(N-K)}\\
\gamma_{1}E_{(N-K)\times K} & \alpha_{1}(q)I_{N-K}+\beta_{1}(q,p)E_{N-K} & \gamma_{2}(q,p)E_{(N-K)\times K} & \alpha_{2}(q)I_{N-K}+\beta_{2}(q,p)E_{N-K}\\
\alpha_{3}(p)I_{K}+\beta_{3}(p)E_{K} & \gamma_{3}E_{K\times(N-K)} & \alpha_{4}(p)I_{K}+\beta_{4}(p,q)E_{K} & \gamma_{4}(p,q)E_{K\times(N-K)}\\
\gamma_{3}E_{(N-K)\times K} & \alpha_{3}(q)I_{N-K}+\beta_{3}(q)E_{N-K} & \gamma_{4}(q,p)E_{(N-K)\times K} & \alpha_{4}(q)I_{N-K}+\beta_{4}(q,p)E_{N-K}
\end{bmatrix},
\end{equation}
where the coefficients $\alpha_{i}$, $\beta_{i}$, $\gamma_{i}$,
$i=1,2,3,4$, are functions defined as follows,
\begin{equation}
\begin{aligned}\alpha_{1}(x)= & -\frac{(N-1)\left[\pi^{2}Nx-4\pi Nx\arccsc\sqrt{N}+4(N-1)\arccsc^{2}\sqrt{N}\right]}{2\pi^{2}Nx\left[N(x-1)+1\right]},\\
\beta_{1}(x,y)= & \frac{1}{2\pi^{2}N^{2}x\left[N(x-1)+1\right](p+q-1)\left[N(p+q)-1\right]}\\
 & \times\left\{ -4(N-1)^{2}\arccsc^{2}\sqrt{N}\left(Nx^{2}-Npq+x-Ny^{2}+Ny\right)\right.\\
 & -4(N-1)\arccsc\sqrt{N}\pi x\left[N(y-1)+1\right]\left[N(p+q)-1\right]\\
 & +\left.\pi^{2}x\left[N(p+q)-1\right]\left[N\left(Nx^{2}+Npq-2Nx+x+1\right)-1\right]\right\} ,\\
\gamma_{1}= & \frac{\left\{ \pi\left[N(p+q)-1\right]-2(N-1)\arccsc\sqrt{N}\right\} ^{2}}{2\pi^{2}N^{2}(p+q-1)\left[N(p+q)-1\right]},
\end{aligned}
\end{equation}
\begin{equation}
\begin{aligned}\alpha_{2}(x)= & \frac{(N-1)\arccsc\sqrt{N}\left[\pi Nx-2(N-1)\arccsc\sqrt{N}\right]}{\pi^{2}Nx\left[N(x-1)+1\right]},\\
\beta_{2}(x,y)= & \frac{1}{2\pi^{2}N^{2}x\left[N(x-1)+1\right](p+q-1)\left[N(p+q)-1\right]}\\
 & \times\left\{ -\pi x\left[N\left(2Nx^{2}+Npq-3Nx+Ny^{2}-Ny+x+3\right)-3\right]\right.\\
 & -4(N-1)^{2}\arccsc^{2}\sqrt{N}\left(Nx^{2}-Npq+x-Ny^{2}+Ny\right)\\
 & \left.+\pi^{2}x\left[N(x-1)+1\right](Nx-1)\left[N(p+q+1)-2\right]\right\} ,\\
\gamma_{2}(x,y)= & \frac{\pi^{2}(Nx-1)\left[N(p+q+1)-2\right]+2(N-1)\arccsc\sqrt{N}\left[3\pi-\pi N(3x+y)+2(N-1)\arccsc\sqrt{N}\right]}{2\pi^{2}N^{2}(p+q-1)\left[N(p+q)-1\right]},
\end{aligned}
\end{equation}
\begin{equation}
\begin{aligned}\begin{aligned}\alpha_{3}(x)= & \frac{(N-1)\arccsc\sqrt{N}\left[\pi Nx-2(N-1)\arccsc\sqrt{N}\right]}{\pi^{2}Nx\left[N(x-1)+1\right]},\\
\beta_{3}(x,y)= & \frac{-(N-1)\arccsc\sqrt{N}}{\pi^{2}N^{2}p\left[N(x-1)+1\right]\left[N(p+q)-1\right](p+q-1)}\\
 & \times\left\{ \pi x\left[N(y-1)+1\right]\left[N(p+q)-1\right]+2(N-1)\arccsc\sqrt{N}\left(Nx^{2}-Npq-x-Ny^{2}+Ny\right)\right\} ,\\
\gamma_{3}= & \frac{(N-1)\arccsc\sqrt{N}\left[-\pi N(p+q)+2(N-1)\arccsc\sqrt{N}+\pi\right]}{\pi^{2}N^{2}(p+q-1)\left[N(p+q)-1\right]},
\end{aligned}
\end{aligned}
\end{equation}
\begin{equation}
\begin{aligned}\begin{aligned}\alpha_{4}(x)= & -\frac{2(N-1)^{2}\arccsc^{2}\sqrt{N}}{\pi^{2}Nx\left[N(x-1)+1\right]},\\
\beta_{4}(x,y)= & \frac{-4(N-1)\arccsc\sqrt{N}\left\{ (N-1)\arccsc\sqrt{N}\left(Nx^{2}-Npq-x-Ny^{2}+Ny\right)+\pi x\left[N(x-1)+1\right](Nx-1)\right\} }{2\pi^{2}N^{2}x\left[N(x-1)+1\right]\left[N(p+q)-1\right](p+q-1)}\\
 & +\frac{y}{2\left[N(p+q)-1\right]},\\
\gamma_{4}(x,y)= & \frac{4(N-1)\arccsc\sqrt{N}\left[-\pi Nx+(N-1)\arccsc\sqrt{N}+\pi\right]}{2\pi^{2}N^{2}(p+q-1)\left[N(p+q)-1\right]}+\frac{y}{2\left[N(p+q)-1\right]}.
\end{aligned}
\end{aligned}
\end{equation}

According to the above results and Eq (\ref{eq:dpsidphi}), the differential
of the initial state $\vert d\psi\rangle$ and the differential of
the reflection axis of the diffusion operator $\vert d\varphi\rangle$
can be derived as
\begin{equation}
\ord{\vert d\psi\rangle}=\ord{\vert d\varphi\rangle}=\frac{2K\sqrt{\frac{N-1}{N}}(Np-1)\arccsc\sqrt{N}}{\pi\left(-2KNp+K+N^{2}p\right)}\begin{bmatrix}(K-N)E_{K\times1}\\
KE_{(N-K)\times1}
\end{bmatrix}d\lambda.\label{eq:first-order-optimal}
\end{equation}
All the other matrices and vectors, such as $\vert d^{2}\psi\rangle$,
$\vert d^{2}\varphi\rangle$, $\vert\frac{dv_{a}}{d\lambda}\rangle$
and $\vert\frac{dv_{b}}{d\lambda}\rangle$, can be obtained accordingly
using the above results by the method in Sec. \ref{subsec:Differential-solution-to}.

Now, if a low failure probability of the quantum search algorithm
$\Delta P$ is allowed, one can minimize the number of oracle calls
with the success probability of the search algorithm fixed by optimizing
the initial state of the quantum system and the reflection axis of
the diffusion operator. The reduction in the number of oracle calls
is characterized by the ratio between the failure probability of the
search algorithm and the squared decrease of the oracle calls, which
is the factor $S$ defined in Eq. (\ref{eq:s}), and the result turns
out to be
\begin{equation}
\ord S=\frac{2Np(Kp-1)}{-2Kp+K/N+Np}.\label{eq:S-1}
\end{equation}

\section{Derivation of higher-order optimal solution\label{sec:Calculation of higher-order expansions}}

In order to obtain a more precise optimal solution to $d\lambda$,
we can plug the first-order (with respect to $d\lambda$) optimal
solution to the initial state $|\psi\rangle$ and the reflection axis
$|\varphi\rangle$ of diffusion (\ref{eq:first-order-optimal}) into
the average success probability $\bar{P}$ (\ref{eq:P}), and expand
$d\bar{P}$ to the third order of $d\lambda$,

\begin{equation}
d\bar{P}=\frac{1}{2}Sd\lambda^{2}+\frac{1}{6}Gd\lambda^{3},\label{eq:dpd}
\end{equation}
where $S$ has already been calculated in the text as $\ord S$ in
Eq. (\ref{eq:S}). The coefficient $G$ can be derived as
\begin{equation}
\ord G=-\frac{12(K-1)K(N-2)Np(Np-1)^{2}(KNp-1)\arccsc\sqrt{N}}{\pi\sqrt{N-1}(-2KNp+K+Np)^{3}}.
\end{equation}
Eq. (\ref{eq:dpd}) will generate a more precise solution to $d\lambda$
in terms of $d\bar{P}$.

To obtain the solution to Eq. (\ref{eq:dpd}), we can use the result
(\ref{eq:dlp}) and extend it the first order of $d\bar{P}$, i.e.,
\begin{equation}
d\lambda=-\sqrt{\frac{2d\bar{P}}{S}}+Bd\bar{P}.\label{eq:dld}
\end{equation}
Then plugging Eq. (\ref{eq:dld}) into Eq. (\ref{eq:dpd}), we get

\begin{equation}
d\bar{P}=\frac{1}{2}S\left(\sqrt{\frac{2d\bar{P}}{S}}+Bd\bar{P}\right)^{2}+\frac{1}{6}G\left(\sqrt{\frac{2d\bar{P}}{S}}+Bd\bar{P}\right)^{3}.\label{eq:eqd}
\end{equation}
Note that although Eq. (\ref{eq:eqd}) involves the third order of
$d\bar{P}$, it is accurate only up to the order $3/2$ of $d\bar{P}$,
as a higher order of $d\bar{P}$ will appear in higher orders of $d\lambda$
which are not included in Eq. (\ref{eq:dpd}). But this equation is
sufficient to give a solution of $d\lambda$ up to the first order
of $d\bar{P}$ as we have only one parameter $B$ to determine in
Eq. (\ref{eq:dld}).

By comparing the terms of $d\bar{P}$ up to the order $3/2$ on both
sides of Eq. (\ref{eq:eqd}), one can derive that

\begin{align}
B=-\frac{(K-1)K(N-2)(Np-1)^{2}\arccsc\sqrt{N}}{\pi\sqrt{N-1}Np(-2KNp+K+Np)(KNp-1)} & ,
\end{align}
Then plugging the result of $B$ back into Eq. (\ref{eq:dld}), one
has a more precise optimal solution to $d\lambda$.

\twocolumngrid

\bibliographystyle{apsrev4-2}
\bibliography{reference309}

\end{document}